\documentclass[preprint]{revtex4}
\usepackage{amssymb}
\usepackage{amsmath}
\usepackage{indentfirst}
\usepackage[mathscr]{eucal}
\usepackage{graphicx}

\begin{document}

\title{
        Heat conductance in nonlinear lattices \\
        at small temperature gradients.
       }

\author{
         T.Yu.~Astakhova, V.N.~Likhachev, G.A.~Vinogradov
        }
\affiliation{
 Emanuel Institute of Biochemical Physics RAS,
 ul.~Kosygina~4, Moscow~119334, Russian Federation
             }

\email{gvin@deom.chph.ras.ru}

\begin{abstract}

This paper proposes a new methodological framework within which
the heat conductance in 1D lattices can be studied. The total
process of heat conductance is separated into two parts where the
first one is the equilibrium process at equal temperatures $T$ of
both ends and the second one -- non-equilibrium with the
temperature $\Delta T$ of one end and zero temperature of the
other. This approach allows significant decrease of computational
time at $\Delta T \to 0$. The threshold temperature $T_{\rm thr}$
is found which scales $T_{\rm thr}(N) \sim N^{-3}$ with the
lattice size $N$ and by convention separates two mechanisms of
heat conductance: phonon mechanism dominates at $T < T_{\rm thr}$
and the soliton contribution increases with temperature at $T >
T_{\rm thr}$. Solitons and breathers are directly visualized in
numerical experiments. The problem of heat conductance in
non-linear lattices in the limit $\Delta T \to 0$ can be reduced
to the heat conductance of harmonic lattice with time-dependent
stochastic rigidities determined by the equilibrium process at
temperature $T$. The detailed analysis is done for the
$\beta$-FPU lattice though main results are valid for
one-dimensional lattices with arbitrary potentials.

\vspace{1 cm}




\end{abstract}

\maketitle

\section{Introduction}%

The problem of heat conductance in low dimensional systems
attracts much attention in last decades (see review \cite{Lep03})
and is motivated by the discovery of quasi-one-dimensional
(nanotubes, nanowires, etc.) and two-dimensional (graphen,
graphan, etc.) systems.

The modern theory of heat conductance was initiated by the
celebrated preprint of E.~Fermi, J.~Pasta and S.~Ulam
\cite{Fer55}, though the primary aim was ``{\it of establishing,
experimentally, the rate of approaching to the equipartition of
energy among the various degrees of freedom}''. Subsequent
investigations demonstrated wide area of consequences in many
physical and mathematical phenomena (see reviews in special
issues of journals CHAOS \cite{Cha05} and Lecture Notes in
Physics \cite{Lec08} devoted to the 50th anniversary of the FPU
preprint).

The dynamical properties of nonlinear systems in microcanonical
ensemble (total energy $E =$ const) were thoroughly analyzed in
most papers. It allows to investigate the dynamics and to get
exact results (soliton \cite{Kru64, Zab65, Dod82} and breather
\cite{Cam96, Sie88, Dau93, Aub94, Mac94} solutions), to analyze
regular and stochastic regimes and to find the corresponding
thresholds. The FPU preprint also initiated the investigations in
the field of ``experimental mathematics'' \cite{Por09} .

About ten decades ago P.~Debye argued that the nonlinearity can
be responsible for the finite value of heat conductance in
insulating materials \cite{Deb14}. But modern analysis shows that
it is not always the case. There are many examples where the
coefficient of heat conductance $\varkappa$ diverges with the
increasing of the system size $L$ as $\varkappa \propto
L^{\alpha}$ where $\alpha > 0$, and $\varkappa \to \infty$ in the
thermodynamic limit ($L \to \infty$). Most of momentum conserving
one-dimensional nonlinear lattices with various types of
nearest-neighbor interactions have this unusual property (see,
e.g., \cite{Lep03, Cas05, Lic08}\,). Moreover, some other
systems, -- two- \cite{Yi_09, But06a, But06b, Fla94} and
three-dimensional lattices \cite{Fla97, Shi08}, polyethylene
chains \cite{Hen09}, carbon nanotubes \cite{Yao05, Yu_05, Mar02,
Min05, Cao04} have analogous property -- diverging heat
conductance with the increasing size of the system.

There were some conjectures explaining the anomalous heat
conductance. Generally speaking, whenever the equilibrium
dynamics of a lattice can be decomposed into that of independent
``modes'' or quasi-particles, the system is expected to behave as
an ideal thermal conductor \cite{Lep05}. Thereby, the existence
of stable nonlinear excitations is expected to yield ballistic
rather than diffusive transport. At low temperatures normal modes
are phonons. At higher temperatures noninteracting ``gas'' of
solitons starts to play more significant role and M.~Toda was the
first, who suggested the possibility of heat transport by
solitons \cite{Tod79}.

Though analytical expressions for solitons can be derived only
for few continuum models described by partial differential
equations, Friesecke and Pego in a series of recent papers
\cite{Fri99, Fri02, Fri04a, Fri04b} made a detailed study of the
existence and stability of solitary wave solutions on discrete
lattices with the Hamiltonian $H = \sum_i \frac12 p_i^2 +
u(y_i)$, where $y_i = x_i - x_{i-1}; \ p_i = \dot x_i$. It has
been proven that the systems with this Hamiltonian and with the
following generic properties of nearest-neighbor interactions:
$u'(0) = 0; \ u''(0) > 0; \ u'''(0) \neq 0$ has a family of
solitary wave solutions which in the small amplitude,
long-wavelength limit have a profile close to that of the KdV
soliton. It was also shown \cite{Hof08} that these solutions are
asymptotically stable. Thus most acceptable point of view on the
origin of anomalous heat conductance in nonlinear lattices is as
follows: phonons are responsible for heat conductance at low
temperatures, and at high temperatures -- solitons
\cite{Li_05,Vil02}.

A set of generic properties were found in a series of papers in
investigation of dynamics of nonlinear lattices, starting from
the celebrated preprint of FPU \cite{Fer55}. And one is an
existence of stochasticity thresholds. The weak stochasticity
threshold is characterized by a specific energy $\mathcal{E}$
below the which the trajectory in the phase space is almost
regular (with near zero Lyapunov exponents) and only small part
of normal modes is excited (it is just the case observed and
analyzed by FPU). The strong stochasticity threshold corresponds
to the value of $\mathcal{E}$ above which energy equipartition
between normal modes is established, and Lyapunov exponents are
positive \cite{Boc70, Pet90, Cas95, Cas97, Gal05, Gal08, Lic08,
Pet05, Zas05}.

A major part of results was obtained using microcanonical
ensemble for isolated systems. Physically more justified is the
usage of canonical ensemble where temperature is kept (on
average) constant by some or other type of heat baths. If the
constant temperature is maintained by the Langevin sources
(random forces with viscous friction), then from the
Fokker-Planck equation the equilibrium Gibbs distribution
immediately follows.

If one starts calculations from arbitrary initial conditions in
canonical ensemble then some time is necessary to achieve the
state of thermodynamic equilibrium. And this stage is not a
trivial one \cite{Lik09}. Firstly and surprisingly, kinetic and
potential energy can relax to equilibrium with different rates;
secondly, obeying the Maxwell velocity distribution function is
not the sufficient condition of achievement the equilibrium. And
the critical stage of achievement the equilibrium (energy
equipartition between normal modes) is the excitation of the most
longwave normal mode. Characteristic times $\tau$ of achievement
the equilibrium can cover very wide range. For instance, there
are well localized excitations in the harmonic lattice with
random masses or, equivalently, random interparticle potentials
(Anderson localization \cite{And58, And61}) where $\tau \gtrsim
10^{300}$ \cite{Lik05}. And this phenomenon is explained by very
weak interaction of localized excitations, if they are centered
near the lattice center, with the heat reservoir located at the
lattice ends.

If to return back to the problem of heat conductance, then one
meets rather confusing experimental and numerical results, e.g.
exponent $\alpha$ in the dependence $\varkappa \propto
N^{\alpha}$ depends on the model under consideration, types of
boundary conditions, used thermostat (Langevin or N\'ose-Hoover
\cite{Nos84, Hoo85}), and also on temperature. For instance,
temperature dependence of heat conductance in carbon nanotubes
decreases as $\varkappa \sim 1/T$ at $T > 10$ K \cite{Mar03};
experimentally is found \cite{Yu_05} that $\varkappa$ also
decreases with the growth of temperature. Different temperature
dependencies $\varkappa$ vs. $T$ were found in 1D nonlinear
lattices. For $\beta$-FPU lattice: $ \varkappa \sim N^{\alpha}
T^{-1}$ at  $T \lesssim 0.1$  and $ \varkappa \sim N^{\alpha}
T^{1/4}$ at $T > 50$  \cite{Aok01} what is usually observed in
insulating crystals. For the interparticle harmonic potentials
and on-site potentials (e.g. Klein-Gordon chains) $\varkappa \sim
T^{-1.35}$, i.e. heat conductance decreases with the growth of
temperature \cite{Aok00}. One more problem is the calculation of
heat conductance at small temperature gradients. Usually these
calculations are very time consuming because of great
fluctuations of heat current and statistical averaging over large
number of MD trajectories is necessary.

The paper organized as follows: in Section II we introduce new
method for the calculation of  the heat conductance which
significantly decreases the computation time and diminishes the
standard error. The method is based on the separation of the
total process of heat conductance into two contributions:
equilibrium and non-equilibrium and the latter one is responsible
for the energy transfer. In the next Section we found that some
quadratic mean values do not exhibit the expected tendency to
reach zero values as the temperature difference $\Delta T \to 0$.
The threshold temperature $T_{\mathrm{thr}}$, separating two
regimes, -- damped and undamped, is revealed. And the
dependencies of $T_{\mathrm{thr}}$ on temperature $T$ and lattice
length $N$ are found. Some modifications of the calculation of
heat conductance in the limit $\Delta T \to 0$ are introduced in
Section V. Direct evidences of the solitons contribution to the
heat conductance are given in the next Section. $\beta$-FPU
lattice is considered as an example.

\newpage


\section{Heat conductance in the $\beta$-FPU lattice}
    \label{sec:heat}

We consider the one-dimensional lattice of $N$ oscillators with
the interaction of nearest neighbors
\begin{equation}
  \label{2-1}
  U = \sum\limits_i u(y_i), \qquad y_i = x_i - x_{i-1}
\end{equation}
and the $\beta$-FPU potential $u(y) = \frac12 y^2 +
\frac{\beta}{4} y^4$ (usually we put $m = \beta = 1$).

Nonequilibrium conditions are necessary for the heat transport
simulation. The most abundant method is the placement of the
lattice into the heat bath with different temperatures of left
$T_+$ and right $T_-$ ends ($T_+ > T_-$). Different types of heat
reservoirs are thoroughly analyzed in \cite{Lep03}. We utilize
the Langevin forces acting on the left $F_+ = \xi_+ - \gamma \dot
x_1$ and right  $F_- = \xi_- - \gamma \dot x_N$ oscillators.
$\left\{ \xi_{\pm} \right\}$ are independent Wiener processes
with zero mean and $\left< \xi_{\pm}(t_1) \, \xi_{\pm}(t_2)
\right> = 2 \gamma T_{\pm} \, \delta(t_1 - t_2)$. $\Delta T =
(T_+ - T_-)$ is the temperature difference. The generalized
Langevin dynamics with a memory kernel and colored noises is also
suggested \cite{Wan07} to correctly account for the effect of the
heat baths.

The following set of stochastic differential equations (SDEs)
\begin{equation}
  \label{2-2}
   \ddot x_i = - \frac{\partial U}{\partial x_i} + \delta_{i1}
   F_+ + \delta_{iN} F_-
\end{equation}
are to be solved to find the heat flux $J$. Then from the Fourier
low $J = - \varkappa \, \nabla T$ the coefficient of heat
conductance is
\begin{equation}
  \label{2-3}
   \varkappa = N {J}/{\Delta T},
\end{equation}
and the problem is to find the heat current $J$. The local heat
flux (from $i$th to $(i+1)$th oscillator) is defined \cite{Zha02}
by
\begin{equation}
  \label{2-4}
  J_{i \to i+1} = \left< F_{i \to i+1}\, \dot x_{i+1}  \right>;
  \qquad F_{i \to i+1} \equiv -U'(x_{i+1} - x_i),
\end{equation}
where $F_{i \to i+1}$ is a shorthand notation for the force
exerted by the $i$th on the $(i+1)$th oscillator and $\left<
\ldots \right>$ is the time averaged. The total heat flux $J$ can
be found as the mean value $J = (N-1)^{-1} \sum_i^{N-1} J_{i \to
i+1}$.


\subsection{Equilibrium and non-equilibrium contributions to the heat conductance}

\begin{figure}
\begin{center}
\includegraphics[width=120mm,angle=0]{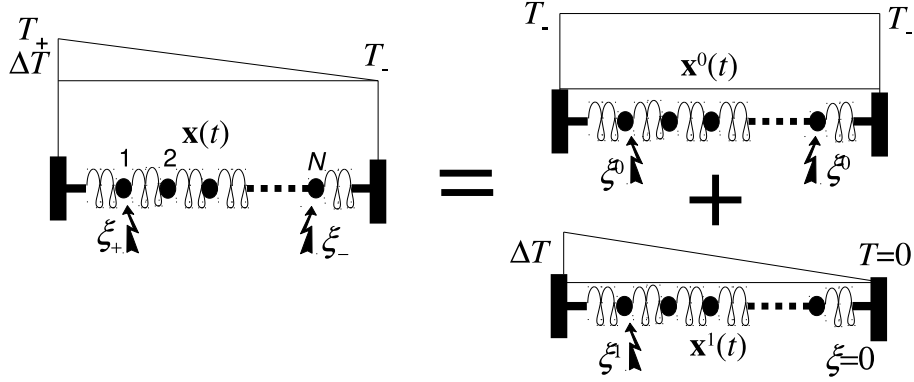}
\end{center}
 \caption{
  \label{fig_01}
  Schematic representation of the splitting of the total
  process ${\bf x}(t)$ into equilibrium ${\bf x}^0(t)$ and
  non-equilibrium ${\bf x}^1(t)$ ones.
 }
\end{figure}

If $T_- \neq 0$ then the process of heat conductance can be
formally separated into two parts: the first one -- equilibrium
process with equal temperatures $T_-$ of both lattice ends; and
second -- nonequilibrium process with temperature $\Delta T$ of
the left lattice end and zero temperature of the right end (see
Fig.~\ref{fig_01}) (by `process' we hereafter assume for brevity
the solution $ {\bf x} (t) =  x_1(t), x_2(t), \ldots , x_N(t); \,
{\bf v} (t) =  v_1(t), v_2(t), \ldots , v_N(t) $ of the
corresponding SDEs).

Namely the second process defines the heat transport realized
against the background of the equilibrium process. Once we
utilize this approach then the Langevin forces in \eqref{2-2} can
be written as $\left\{ \xi_+ \right\} = \left\{ \xi^0 \right\} +
\left\{ \xi^1 \right\} $ and $ \left\{ \xi_- \right\} =  \left\{
\xi^0  \right\} $ for the left and right lattice ends,
correspondingly; superscripts `0' and `1' refer to equilibrium
and nonequilibrium processes. Then the total dynamical process
${\bf x}(t) $ can be represented as the sum of two processes
\begin{equation}
  \label{2-5}
    {\bf x}(t) = {\bf x}^0(t) + {\bf x}^1(t),
\end{equation}
where ${\bf x}^0(t)$ is the equilibrium (Gibbs's) process at
temperature $T_-$, and ${\bf x}^1(t)$ -- nonequilibrium,
responsible for the energy transport, process. Then the Langevin
dynamics is
\begin{equation}
  \label{2-6}
   \ddot x_i^0 =  - \frac{\partial U^0}{\partial x_i} +
                     \delta_{i1} (\xi^0 - \gamma\dot x^0_1) +
                     \delta_{iN} (\xi^0 - \gamma\dot x^0_{N}),
\end{equation}
\begin{equation}
  \label{2-7}
   \ddot x_i^1 =   - \left[
                  \frac{\partial U}{\partial x_i} -
                  \frac{\partial U^0}{\partial x_i}
                     \right] +
                  \delta_{i1} (\xi^1 - \gamma\dot x^1_1) +
                  \delta_{iN} (- \gamma\dot x^1_N),
\end{equation}
and the sum of equations \eqref{2-6} and \eqref{2-7} is virtually
identical to the parent equation \eqref{2-2}. Random values
$\left\{ \xi^0 \right\}$ and $\left\{ \xi^1 \right\}$ obey the
identities $\left< \xi^0(t_1) \xi^0(t_1) \right> = 2 \gamma T_-
\delta(t_1 - t_2)$ and $\left< \xi^1(t_1) \xi^1(t_1) \right> = 2
\gamma \Delta T \delta(t_1 - t_2)$; $U^0$ is the total energy
\eqref{2-1} where the arguments $ x_1(t), x_2(t), \ldots, x_N(t)
$ of the total process are substituted to the coordinates of the
equilibrium process $x_1^0(t), x_2^0(t), \ldots, x_N^0(t)$.
Expression in the square brackets in \eqref{2-7} is the
difference of forces acting on the $i$th particle from the total
process ${\bf x}(t)$ and equilibrium process ${\bf x}^0(t)$.  It
is significant to note that this force is the random value, and
the process ${\bf x}^1(t)$ (heat transport) is realized in the
lattice with {\it time-dependent random potentials}. The problem
of heat conductance in the random time-independent potentials was
analyzed in \cite{Joh08}

Equation \eqref{2-6} describes the system embedded in the heat
reservoir at temperature $T_-$. And ${\bf x}^0(t)$ is the
stationary equilibrium process described by the canonical Gibbs
distribution (equilibrium thermodynamics of the $\beta$-FPU
lattice in the canonical ensemble was considered in
\cite{Lik09}).

Process  ${\bf x}^1(t)$ is responsible for the heat transport and
the Wiener's process $\left\{ \xi^1(t) \right\}$ on the left
lattice end defines small temperature $\Delta T$. Right lattice
end has zero temperature. An expression for the local heat flux
is
\begin{equation}
  \label{2-8}
 J_{i \to i+1} = \left<
        F_{i \to i+1}({\bf x}) \, \dot x_{i+1} - F_{i \to i+1}({\bf x}^0) \, \dot
        x_{i+1}^0 \right>,
\end{equation}
and the equilibrium process ${\bf x}^0$ does not transfer energy:
$\left<  F_{i \to i+1}({\bf x}^0) \, \dot x^0_{i+1} \right>
\equiv 0$.

One of the goals of the present paper is the calculation of heat
conductance at small temperature gradients. Usually these
calculations are realized by solving SDEs \eqref{2-2} and are
very time consuming because of great fluctuations of heat current
(below we show that the time of computation increases $\propto
\left( \Delta T \right)^{-2}$ if the accuracy of calculations is
predetermined).

\begin{figure}
\begin{center}
\includegraphics[width=120mm,angle=0]{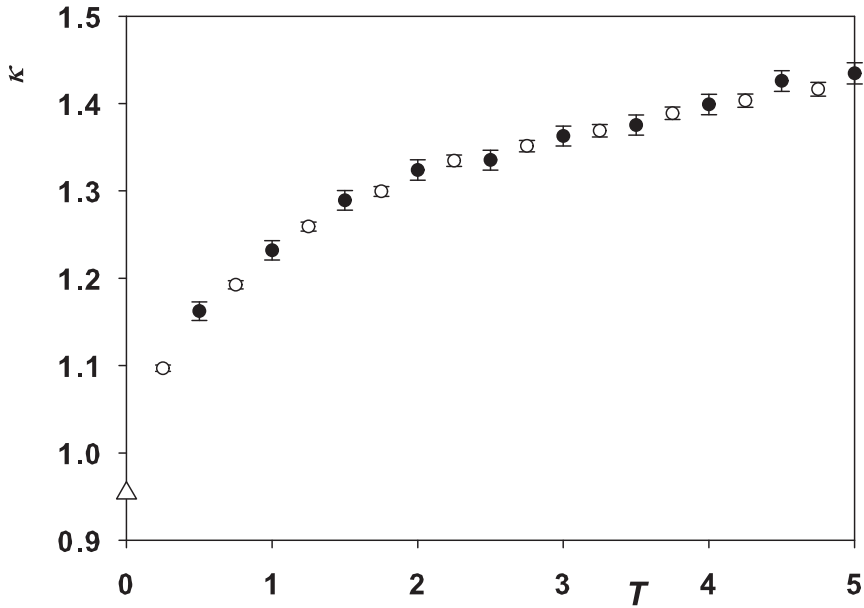}
\end{center}
 \caption{
  \label{fig_02}
  Coefficient of heat conductance vs. temperature for the lattice of $N=5$
  oscillators. Filled circles: solution of standard SDEs \eqref{2-2};
  empty circles: SDEs
  \eqref{2-6}-\eqref{2-7}. Averaging over $100$ MD trajectories
  $10^4$ time units (t.u.) each. $T_- = 0{.}2$,
  $\Delta T = 0{.}01 T_-$, $\gamma = 1$. Triangle up at
  $T = 0$ is the exact value in the harmonic approximation ($\beta = 0$).
         }
\end{figure}

The comparison of two approaches (solving of standard SDEs
\eqref{2-2} and \eqref{2-6}-\eqref{2-7}) is shown in
Fig.~\ref{fig_02} and results coincide with very good accuracy.
Note, that most of results in this paper are presented for the
number of oscillators $N = 5$ in the lattice. It may appear that
this value is too small. For instance, the best estimate so far
required simulations of up $\gtrsim 10^4$ particles and $\gtrsim
10^8$ integration steps plus ensemble averaging \cite{Lep03}. But
our results are aimed at founding some basic issues where number
of particles is less essential. Lattices with larger number of
oscillators were tested where necessary.

\begin{figure}
\begin{center}
\includegraphics[width=120mm,angle=0]{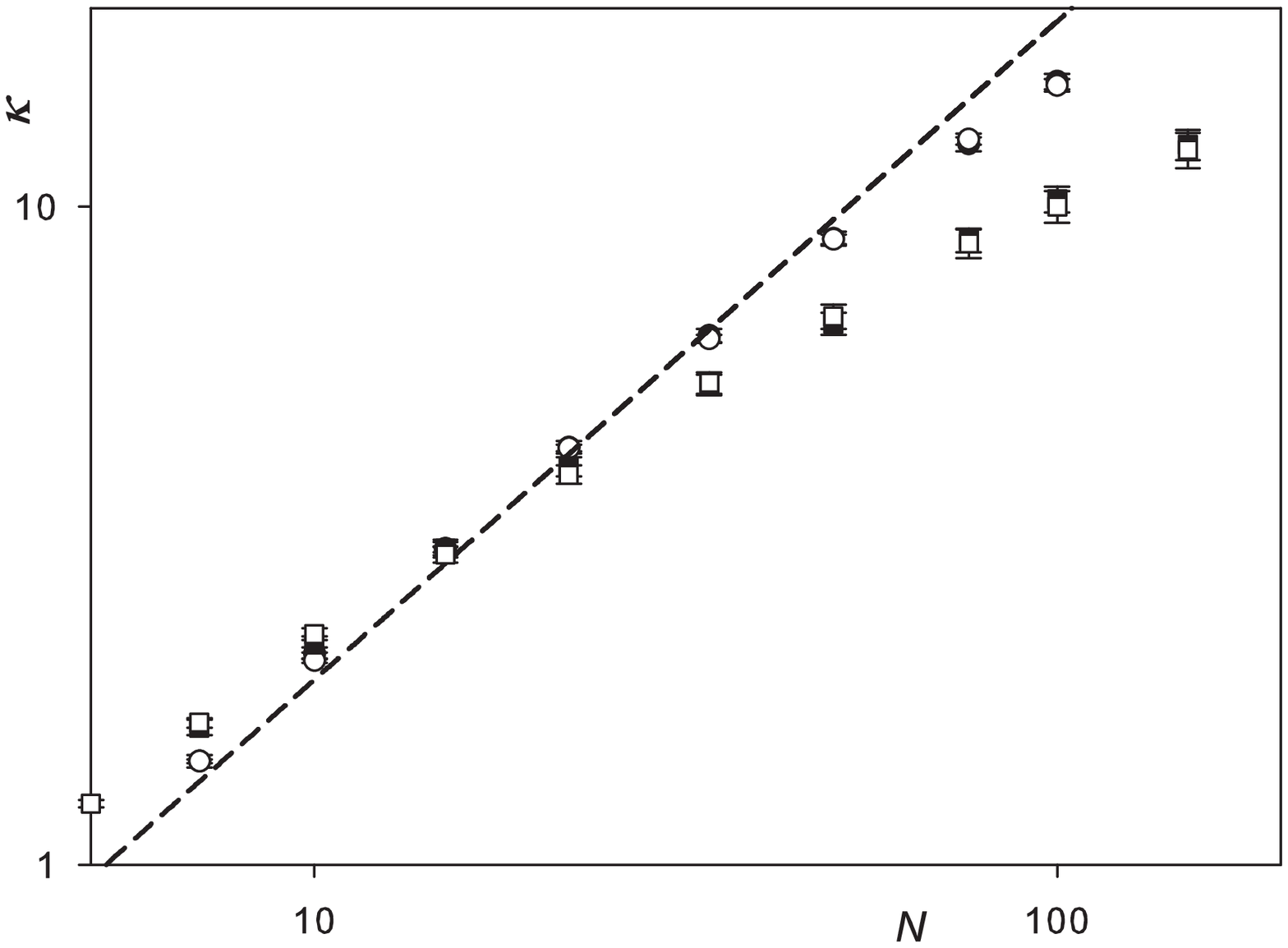}
\end{center}
 \caption{
  \label{fig_3}
  Coefficient of heat conductance for the $\beta$-FPU lattice for
  $N = 7-150$ oscillators. Squares: $T_-=1$, circles:
  $T_-=0{.}1$. Filled symbols -- results obtained by the solution
  of standard SDEs \eqref{2-2}, empty symbols -- SDEs
  \eqref{2-6}-\eqref{2-7}. Averaging over 200 MD trajectories
  $3\,10^4$ t.u. $\Delta T = 0{.}01T$, $\gamma = 1$. Dashed line -- harmonic
  approximation at $T_- \to 0$.
 }
\end{figure}

The dependence of heat conductance on the particles number $N$ is
shown in Fig.~\ref{fig_3} at two value of temperature $T_-$.
Inharmonicity becomes negligible in the limit $T_- \to 0$ and the
analytical solution of the heat conductance for the harmonic
lattice is given in \cite{Rie67}.

There should be solved twice as large SDEs
\eqref{2-6}-\eqref{2-7} in suggested approach as that in standard
scheme \eqref{2-2}, and this the price which is paid for the
facility with using small temperature gradients. As one would
expect, the accuracy of the suggested approach is higher
(provided that all computational terms and conditions are
identical). The comparison of accuracies is given in
Appendix~\ref{Accur_comp}


\section{Strange behavior of process ${\bf  x}^1(t)$ at high temperatures}

Usually the temperature difference $\Delta T \sim (0{.}01 -
0{.}1) T_-$ at the lattice ends is an appropriate choice. Then
the Fourier law $J \propto \Delta T$ (at fixed $N$) is valid with
good accuracy. Actually, the corrections to the heat current are
of the order $(\Delta T)^3$ as the current is the odd function of
the temperature difference, and this ensures the reasonable
accuracy of the linear approximation.

Now we concentrate our efforts on the elucidating the heat
conductance dependence via temperature of the background process
${\bf x}^0(t)$. Langevin forces $\left\{ \xi^1 \right\}$, which
provide temperature $\Delta T$, are of the order $ \xi^1 \sim
\sqrt{\Delta T}$ (as $\left< \xi^1(t_1) \xi^1(t_2) \right> \sim
\Delta T$). And one can expect that process ${\bf x}^1(t)$ should
have the same order ${\bf x}^1(t) \sim \sqrt{\Delta T}$ because
equation \eqref{2-7} becomes linear in the limit $\Delta T \to 0$
when $\xi^1 \to 0$. Thus any quadratic mean values should be of
the order $\sim \Delta T$.

\begin{figure}
\begin{center}
\includegraphics[width=120mm,angle=0]{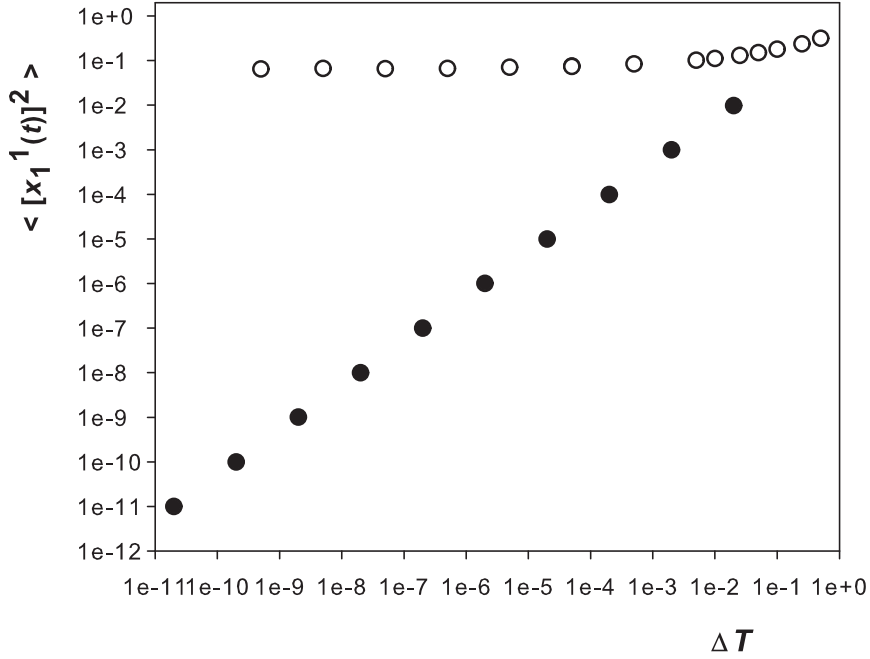}
\end{center}
 \caption{
  \label{fig_4}
  Dependence of the mean value of quadratic form $\left< (x_1^1)^2
  \right>$ on the temperature difference $\Delta T$. Filled circles:
  $T = 0{.}2$ ,  empty circles: $T=5$. Asymptotic value $\left< (x_1^1)^2
  \right>_{\Delta T \to 0} = 0{.}064$ at $T = 5$ (coefficient of linear
  regression $0{.}9993$). Averaged over 100 MD trajectories
  $10^4$ t.u. each. $N=5$. The range of $\Delta T$:
  $10^{-11} \leq \Delta T \leq 2 \! \cdot \! 10^{-1}$
 }
\end{figure}

Two temperatures of the background process $T_-$ were tested:
$T_1 = 0{.}2$ and $T_2 = 5$ (from here we omit subindex `--' for
brevity). Mean value $\left< \, [x_1^1(t)]^2 \right>$ was
analyzed as an example and results are shown in Fig.~\ref{fig_4}
(fully identical properties have all quadratic values
(correlators) of the types $\left< \, x_i^1(t) x_j^1(t) \right>,
\ \left< \, \dot x_i^1(t) \dot x_j^1(t) \right> \ \left< \,
x_i^1(t) \, \dot x_j^1(t) \right>$). As one expects, the
quadratic form $\left< \, [x_1^1(t)]^2 \right>$ linearly depends
on $\Delta T$: $\left< \, [x_1^1(t)]^2 \right> \sim \Delta T$ at
$T_1 = 0{.}2$. But the case is quite different at $T_2 = 5$: mean
value $\left< \, [x_1^1(t)]^2 \right>$ tends to a stationary
value $0{.}064$ in the limit $\Delta T \to 0$. It means that
there exists some undamped stationary process ${\bf x}^1(t)$ at
high temperatures $T$ even in the limit $\Delta T \to 0$. These
results also can imply an existence of a threshold temperature
$T_{\rm thr}$ separating two regimes -- damped at low
temperatures and undamped at high temperatures.


\section{The threshold temperature}

Any process ${\bf x}^1(t)$ damps out at low temperatures and
flattens out to a stationary value at higher temperatures even in
the limit $\Delta T \to 0$, and the temperature $T$ of process
${\bf x}^0(t)$ determines the different damping rates. And an
illustrative process $\widetilde{\bf x}^1(t)$ was analyzed to
determine an existence of a threshold temperature and its value
('tilde' marks the process ${\bf x}^1(t)$ at $\Delta T = 0$ to
avoid confusions).

Process $\widetilde{\bf x}^1(t)$ can be exited in some or other
manner. Usually ${\widetilde x}_i^1$ and ${\widetilde v}_i^1$ get
random values in such a way that $\frac12 \sum_i [{\widetilde
x}_i^1(t=0)]^2 = \frac12 \sum_i [ {\widetilde v}_i^1(t=0)]^2 =
0{.}5$. The particular choice of initial conditions does not
influence the final results.

Stochastic differential equation for the process ${\bf \widetilde
x}^1(t)$ are
\begin{equation}
  \label{2-10}
   \ddot {\widetilde x}_i^1 =   - \left[
                  \frac{\partial  U } {\partial x_i} -
                  \frac{\partial  U^0 }{\partial x_i}
                     \right] -
                  \delta_{i1}\dot {\widetilde x}^1_1 -
                  \delta_{iN} \dot {\widetilde x}^1_{N};
                  \qquad (\gamma = 1),
\end{equation}
with random forces determined by the difference of processes
${\bf x}(t)$ and ${\bf x}^0(t)$, and damping at the extreme left
and right oscillators; $U$ and $U^0$ are potential energies with
coordinates ${\bf x}(t)$ and ${\bf x}^0(t)$, correspondingly.
Stochastic dynamics  \eqref{2-10} is implicitly ruled out by the
temperature $T$ of process ${\bf x}^0(t)$.


\subsection{Two methods to find $T_{\rm thr}$}

We consider the case of small temperature $T$ when process ${\bf
\widetilde x}^1(t)$ is damped out. The damping is determined by
the viscous friction of left $(-\dot {\widetilde x}^1_1)$ and
right $(-\dot {\widetilde x}^1_{N})$ oscillators in \eqref{2-10}.
Gradually increasing the temperature we find its threshold value
when process ${\bf \widetilde x}^1(t)$ becomes undamped.

\begin{figure}
!\begin{center}
\includegraphics[width = 80mm, angle=0]{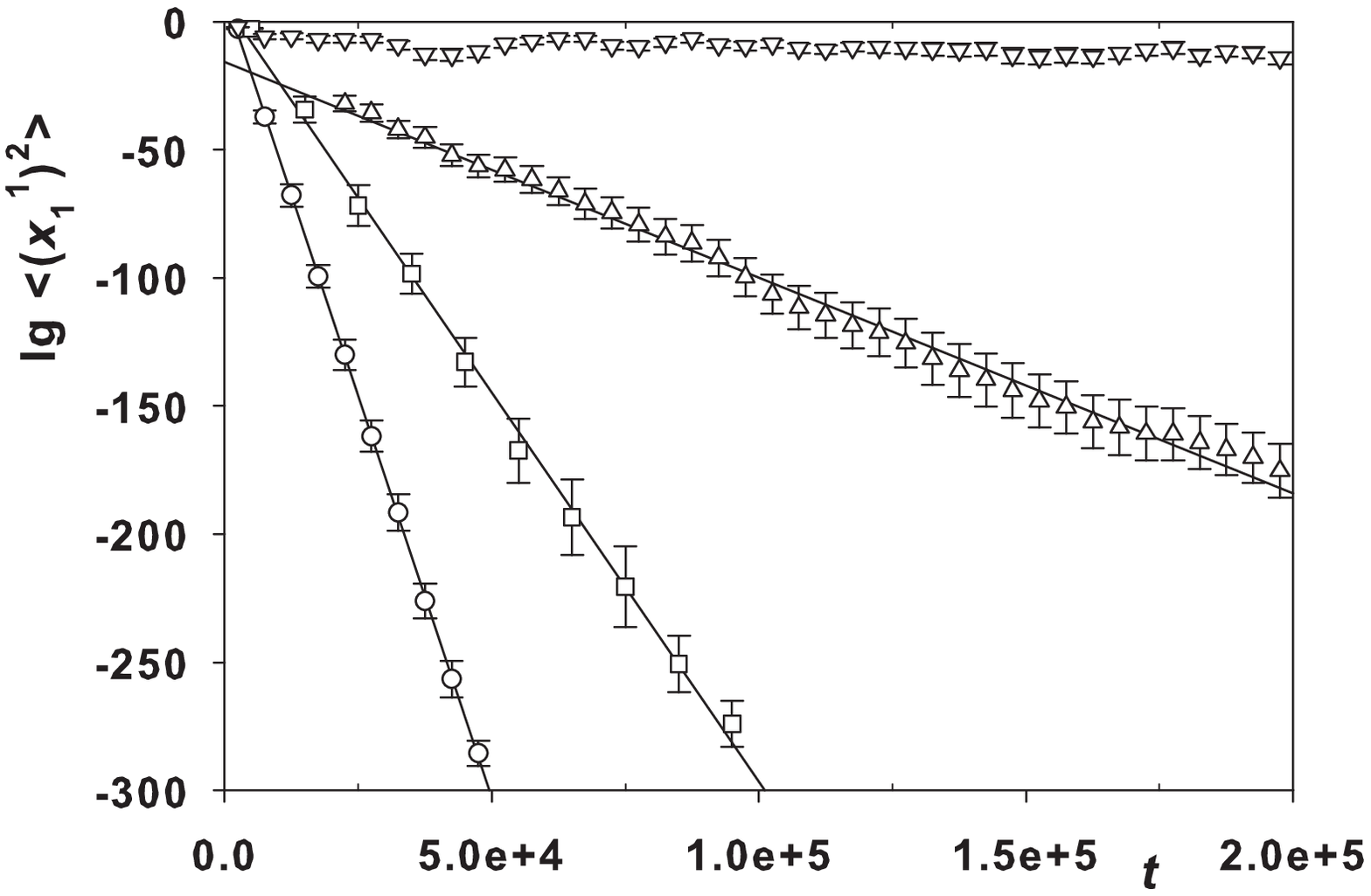}
\includegraphics[width = 75mm, angle=0]{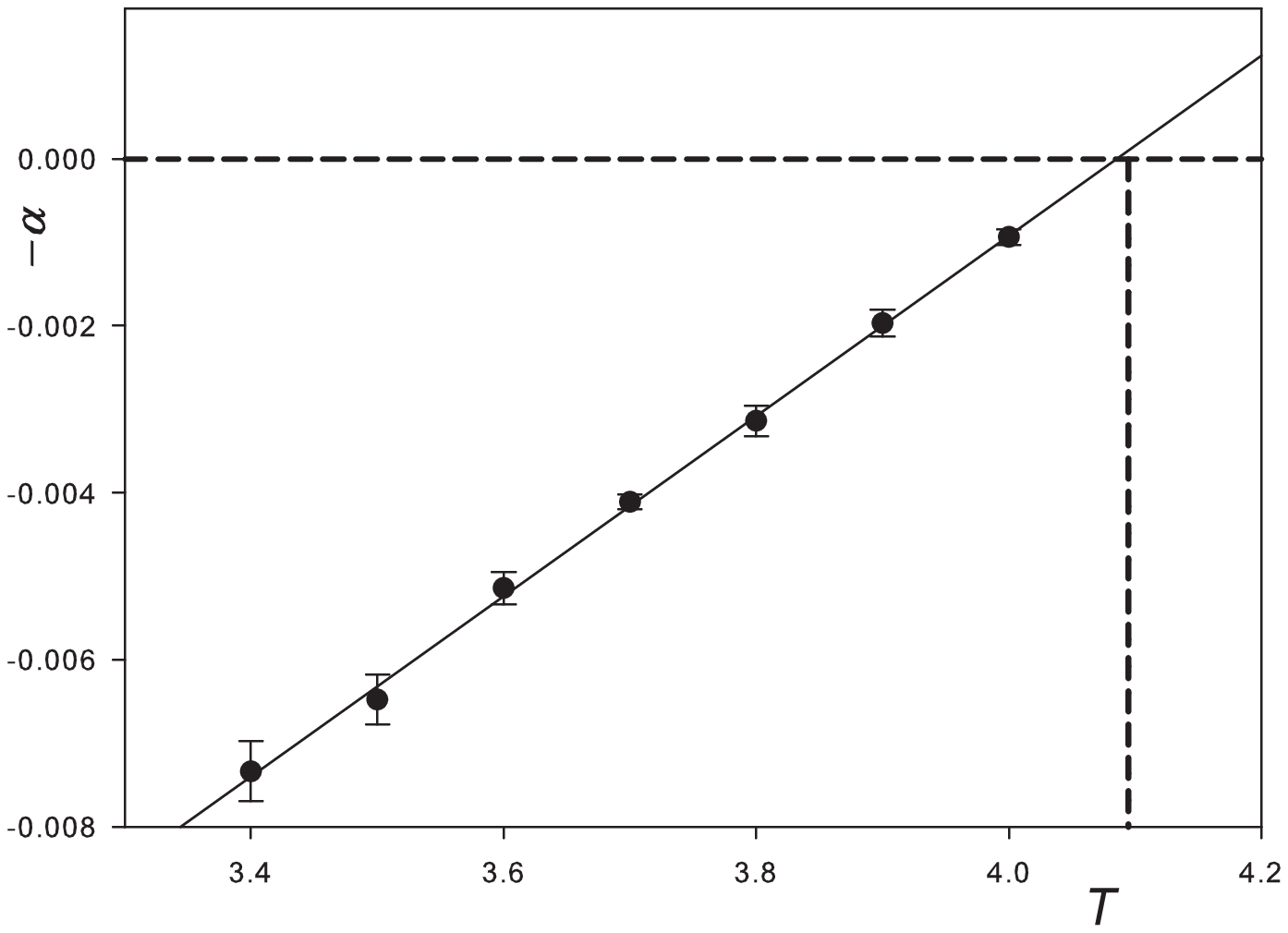}
!\end{center}
 \caption{
  \label{fig_5}
  a) Exponential damping of process ${\bf \widetilde x}^1(t)$ at
  different temperatures: $T=3{.}5$ (circles), $T=3{.}8$ (squares),
  $T=4{.}0$ (triangles up), $T=4{.}2$ (triangles down). Solid
  lines -- linear regression. Averaging time  $\sim \! 5\,000-10\,000$
  t.u. 20 trajectories ${\bf x}^0$ were used to estimate the standard
  error. b) Damping coefficient $(-\alpha)$ as the function of temperature
  $T$ of process ${\bf x}^0(t)$. Damping stops $(\alpha = 0)$ at
  $T_{\rm thr} \simeq 4{.}07$. $N=5$.
 }
\end{figure}

The damping of mean squared displacement of the first oscillator
${\widetilde x}_1^1(t)$ was calculated. It was found that this
process exponentially decays $\left< \, [\widetilde x_1^1(t)]^2
\right> \propto \exp(-\alpha t)$ and $\alpha$ depends on $T$ (see
Fig.~\ref{fig_5}a). One can see that the damping stops in the
range $4{.}0 < T < 4{.}2$.
The dependence of coefficient $\alpha$ on the temperature $T$ of
process ${\bf x}^1(t)$ is shown in Fig.~\ref{fig_5}b and $T_{\rm
thr} \simeq 4{.}07$ at $\alpha = 0$.

\begin{figure}
\begin{center}
\includegraphics[width=100mm,angle=0]{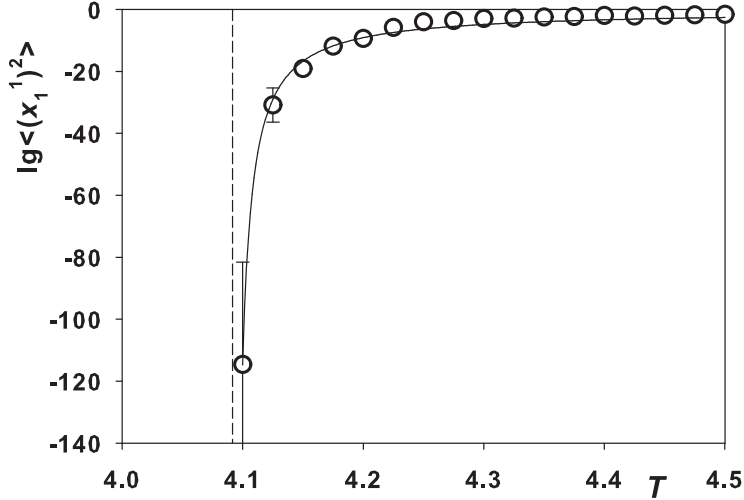}
\end{center}
 \caption{
  \label{fig_6}
  Stationary values $\left< (\widetilde x_1^1)^2 \right>$ at
  temperatures higher then $T_{\rm thr}$ in log-linear coordinates.
  Time of averaging $~10^6$ t.u. The temperature dependence was
  approximated by the function
  $\left< [\widetilde x_1^1(t)]^2 \right>
  \sim \exp[-b /(T-T_{\rm thr})]$ at $T > T_{\rm thr}$ (solid line).
  $T_{\rm thr} \simeq 4{.}09$. $N=5$.
 }
\end{figure}

Now we find the threshold temperature going ``from up to down'',
going from higher temperatures. At high temperatures there exists
the stationary process outcoming from random forces $\Phi$
(see~\eqref{2-7} -- expression in square brackets). Process ${\bf
\widetilde x}^1(t)$ decreases in the sense that all quadratic
mean values tend to zero as temperatures approaches $T_{\rm
thr}$. When the threshold temperature reaches it threshold value,
process ${\bf \tilde x}^1(t)$ disappears (see Fig.~\ref{fig_6}).
The found threshold temperature is $T_{\rm thr} \simeq 4{.}09$.


\subsection{Time-resolved dynamics of process ${\bf x}^1(t)$ }

\begin{figure}
\begin{center}
\includegraphics[width=77mm,angle=0]{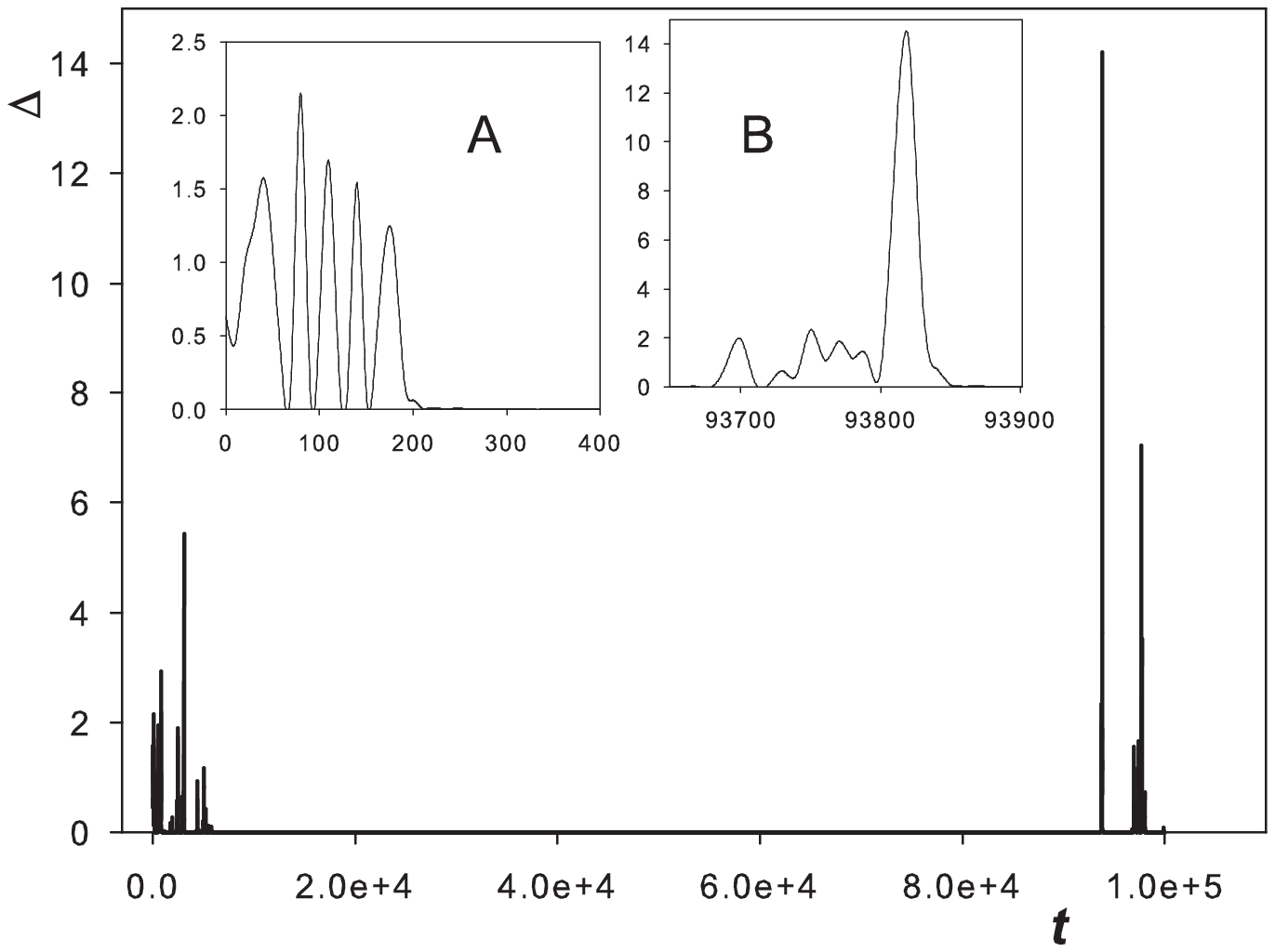}
\includegraphics[width=80mm,angle=0]{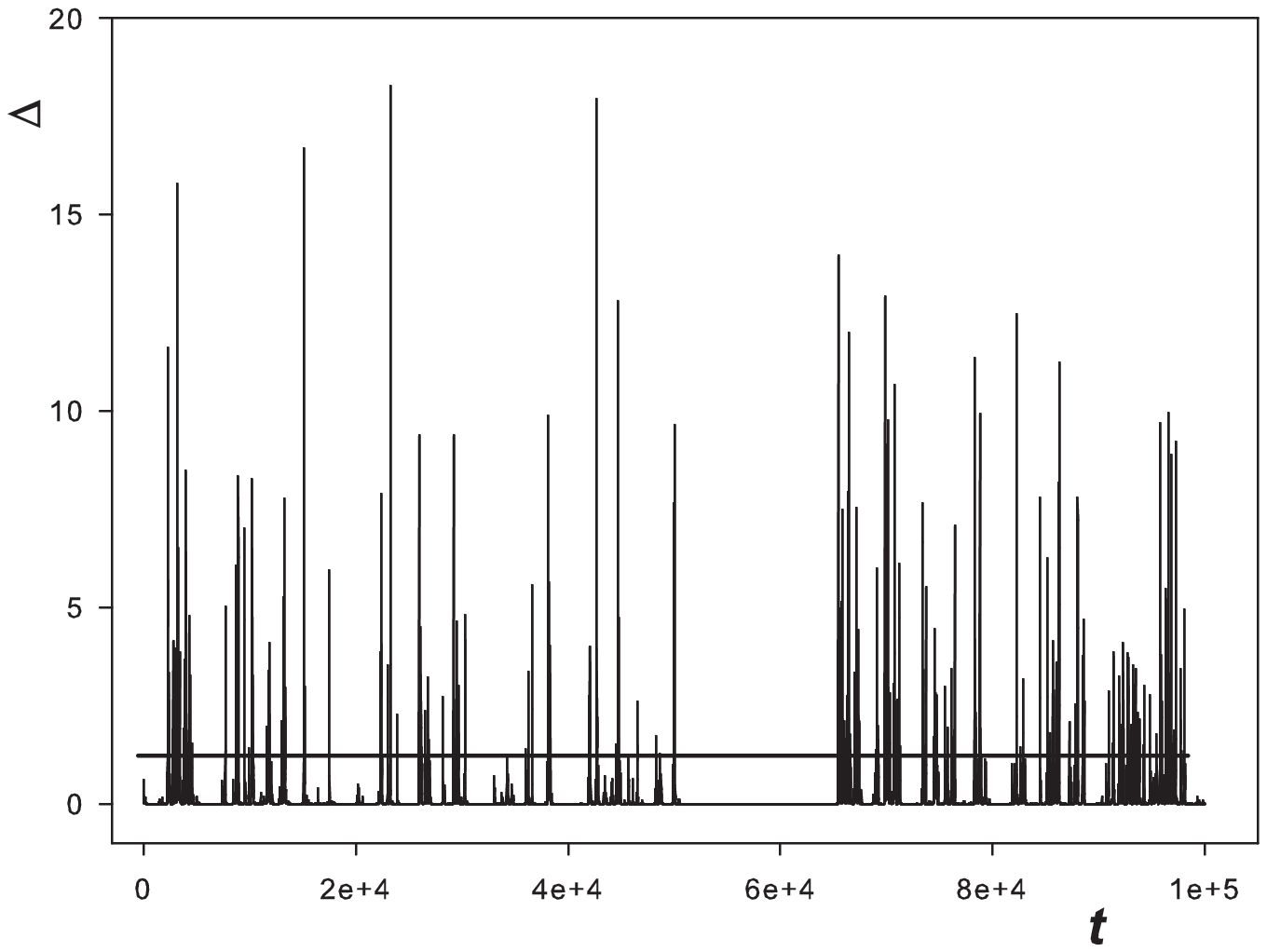}
\includegraphics[width=85mm,angle=0]{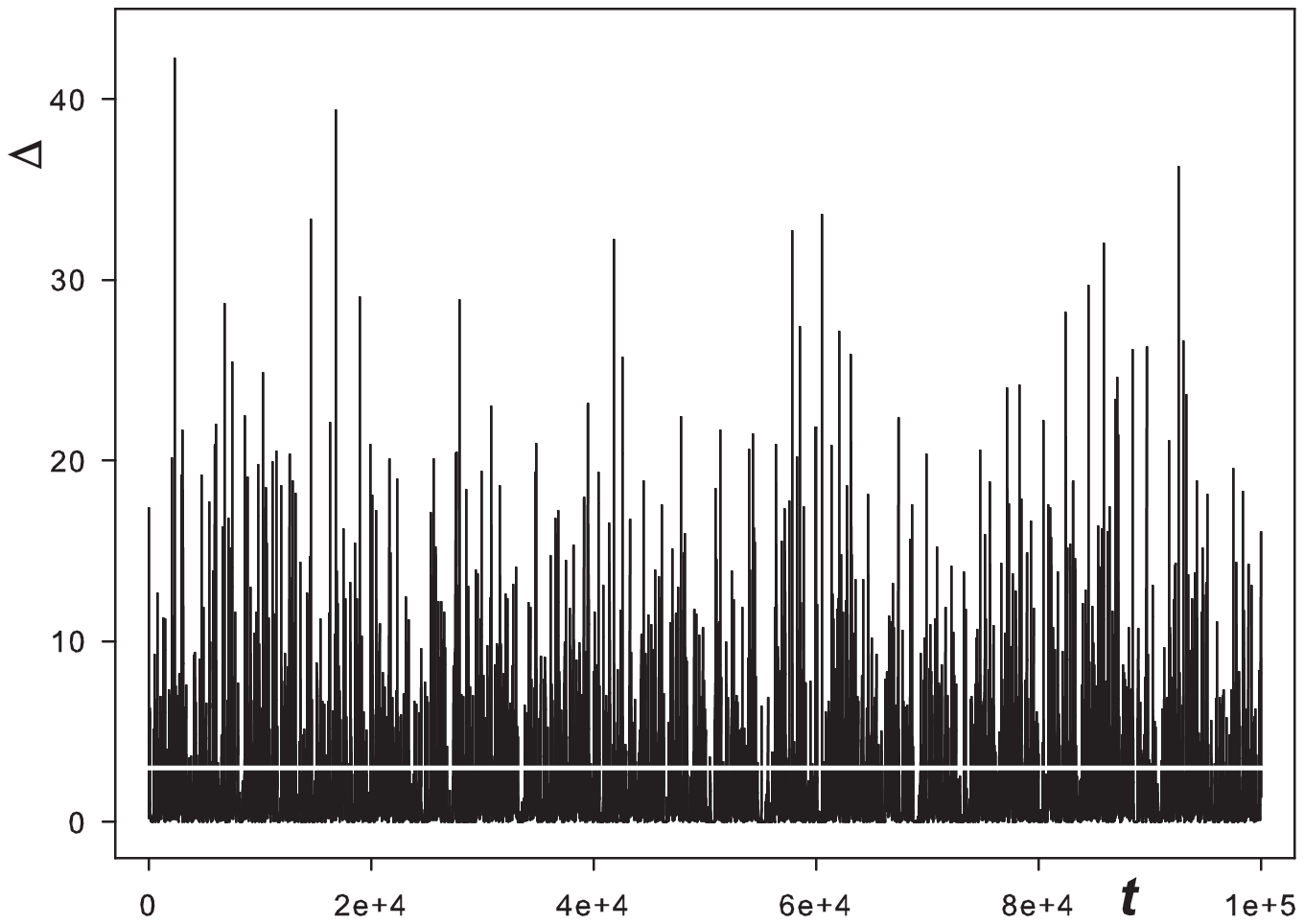}
\end{center}
 \caption{
  \label{fig_7}
Dependence of ${\Delta}(t)$ versus time at different temperatures
of process ${\bf x}^0(t)$. a)  $T=3{.}9$; b) $T=4{.}3$; c)  $T =
7{.}0$. $N=5$, integration step $h = 0{.}01$. Oscillations at
small times (insert `A' to panel a) decay on average
exponentially. Detailed shape of ``excitation'' at $t \approx
93\,800$ is shown in insert `B'. Mean values $\left. \left<
\Delta(t) \right> \right|_{t=0}^{t=10^5}$ are shown in horizontal
solid lines.
 }
\end{figure}

To elucidate the reasons of strange dynamics of process ${\bf
x}^1(t)$ at high temperatures $T$ we analyzed it more thoroughly.
As above, ${\Delta}(t) = [\widetilde x_1^1(t)]^2$ was calculated
but without averaging over time. Results are shown in
Fig.~\ref{fig_7} for three temperatures $T$ of process ${\bf
x}^0$.
One can see that ${\Delta}(t)$ behaves highly irregular. And
numbers and heights of observed peaks increases with the growth
of temperature until it becomes chaotic at high $T$. The mean
values $\left< {\Delta}(T) \right>_{\tau}$ at different
temperatures $T$ averaged over time interval $\tau = 10^5$ t.u.
increase with temperature. Mean values $\left< \Delta(t)
\right>$, shown in horizontal solid lines, are nothing else than
the stationary values calculated above. It was specially checked
out that the dynamics observed in Fig.~\ref{fig_7} is not due to
numerical artifacts.


\subsection{Heat conductance at small temperature gradients}

Our main concern is the computation of heat conductance at small
temperature gradients. With this in mind we analyze an expression
for the heat current in more details. And this analysis can also
shed some light upon the problem why process ${\bf x}^1(t)$
behaves in such strange manner. Remind an expression for force in
the heat current: $F_{i \to i+1} = -u'_{x_{i+1}}(x_{i+1} -
x_{i})$ is the force acting on the $(i+1)$th oscillator from left
to right, and the derivative of the potential energy $u(x_{i+1} -
x_{i})$ between oscillators is taken with respect to $x_{i+1}$.
Then the expression for the local heat current \eqref{2-8} can be
rewritten as
\begin{equation}
  \label{2-9}
  \begin{split}
 {} &   J_{i \to i+1} = \left< \left[ F_{i \to i+1}({\bf x}) -
                                      F_{i \to i+1}({\bf x}^0)  \right]
    \dot x_{i+1}^1 \right> = \\
 {} &  \left<  \left[ -(x_{i+1} - x_i) -(x_{i+1} - x_i)^3 +
                       (x^0_{i+1} - x^0_i) + (x^0_{i+1} - x^0_i)^3   \right]
                       \dot x_{i+1}^1
        \right>,
  \end{split}
\end{equation}
where $\dot x_{i+1}^1$ -- velocity of $(i+1)$th oscillator in
process ${\bf x}^1$. Difference of forces (in square brackets in
the second line) is the polynomial of the third degree in the
square root of temperature difference $\sqrt{\Delta T}$ (process
$\bf x^1$ is of the order of $\sqrt{\Delta T}$ as discussed
above), and taking into account that the velocity $\dot
x_{i+1}^1$ is also of the order of $\sqrt{\Delta T}$, it is the
polynomial of the forth degree in $\sqrt{\Delta T}$. But the
coefficient of heat conductance is determined by the relation
$J/\Delta T$ therefor terms of the third and forth orders can be
neglected at small values of $\Delta T$. Then \eqref{2-9} is
simplified to
\begin{equation}
  \label{2-91}
     J_{i \to i+1} = \left< - \left( x_{i+1}^1 - x_i^1 \right)
     \left[ 1 + 3 \, \left( x_{i+1}^0 - x_i^0 \right)^2   \right]
     \dot x_{i+1}^1 \right>.
\end{equation}

It is significant that the total heat current in \eqref{2-9} at
large temperature $T$ and small temperature gradient $\Delta T$
is the difference of {\it finite} terms from processes ${\bf
x}(t)$ and ${\bf x^0}(t)$, vanishing in the limit $\Delta T\to
0$. And it is the reason why direct MD simulation is highly
inefficient in this case and gives very large fluctuations. But,
as will be shown below, there exists an efficient method to
overcome this difficulty.

The behavior of process ${\bf x}^1(t)$ is explained by the fact
that it is determined not only by random Langevin forces $\xi^1
\sim \sqrt{ \Delta T}$, but also (and more essentially) by {\it
time-dependent random} forces $\Phi_i = \left[ {\partial
U}/{\partial x_i} - {\partial U^0}/{\partial x_i} \right]$ (see
\eqref{2-7}). The plateau for the correlator $\left< [x^1_1(t)]^2
\right>$ equal to $0{.}064$ at $\Delta T \to 0$ is determined
exclusively by random forces $\Phi_i$ from stationary process
${\bf x}^0$ (an illustrative example of one variable is
considered in Appendix~\ref{One_var}). Thus, the dynamical
process ${\bf x}^1(t)$ becomes the stationary one, determined by
the background process ${\bf x}^0(t)$ at high temperatures.


\section{Threshold temperature in the limit $\Delta T \to 0$}

In this section process ${\bf x}^1(t)$ is considered at an
arbitrary temperature $T$ and in the limit $\Delta T \to 0$.
Remind that process ${\bf x}^1(t) \sim \xi^1 \sim \sqrt{\Delta
T}$ is completely suppressed at $T < T_{\rm thr}$. To realize the
limiting transition $\Delta T \to 0$ in \eqref{2-7} it is
convenient to divide both sides by $\sqrt{\Delta T}$. Then new
coordinates are ${\bf y}(t) = {\bf x}^1(t)/\sqrt{\Delta T}$. It
is also convenient to introduce normalized to unity random force
$\theta = \xi^1/\sqrt{\Delta T}$. Then the linear equation for
${\bf y}(t)$ can be obtained as quadratic and cubic forces can be
neglected (see \eqref{2-9}). The corresponding equation for ${\bf
y}(t)$ can be derived if to rearrange one term in potential
energy \eqref{2-1} keeping in mind that $x_i = x_i^0 + x_i^1$.
Then $u(x_i - x_{i-1}) = \frac12 [(x_i^0 - x_{i-1}^0) + (x_i^1 -
x_{i-1}^1)]^2 + \frac14 [(x_i^0 - x_{i-1}^0) + (x_i^1 -
x_{i-1}^1)]^4$.

Transforming variables to ${\bf y}$ and retaining terms quadratic
in ${\bf y}$, one can get the potential energy in the form
\begin{equation}
  \label{2-11}
  u = \frac12 \sum_i g_i(t) \, (y_i - y_{i-1})^2,
  \qquad g_i(t) = 1 + 3 \, [x_i^0(t) - x_{i-1}^0(t)]^2 \,,
\end{equation}
where  $g_i(t)$ are {\it time-dependent random} coefficients of
rigidity determined by the dynamical process ${\bf x}^0(t)$. It
is illuminating to note that the problem of heat conductance can
be reduced to the {\it quadratic} potential energy in the limit
$\Delta T \to 0$. Corresponding SDEs have Langevin source with
unit temperature at the left oscillator and zero temperature at
the right oscillator:
\begin{equation}
  \label{2-12}
  \ddot y_i = - g_i (y_i - y_{i-1}) + g_{i+1} (y_{i+1} - y_{i}) +
  \delta_{i1} (\theta - \dot y_1) - \delta_{i N} \dot y_{N}  \,.
\end{equation}
It should be also noted that if the 1D lattice with an arbitrary
interaction (Morse, Toda, LJ, etc) is analyzed then the
corresponding equation will be the same, and random rigidities
are $g_i=U''(x^0_i - x^0_{i-1}$) where $U$ is some or other type
of potential energy. Equation for arbitrary systems (with
arbitrary neighbor radius of interaction) can be also written in
the general form as
\begin{equation}
  \label{2-13}
  \ddot y_i = - \sum_{j=1}^M \Lambda_{ij}^0 \, y_j  +
  \delta_{i1} (\theta - \dot y_1) - \delta_{i N} \dot y_{N}
  \,,
\end{equation}
where $\Lambda_{ij}^0$ -- matrix of second derivatives of
potential energy depending on ${\bf x}^0$, and $M$ is the number
of neighbors. Equation \eqref{2-13} is valid for arbitrary
systems.

\begin{figure}
\begin{center}
\includegraphics[width=100mm,angle=0]{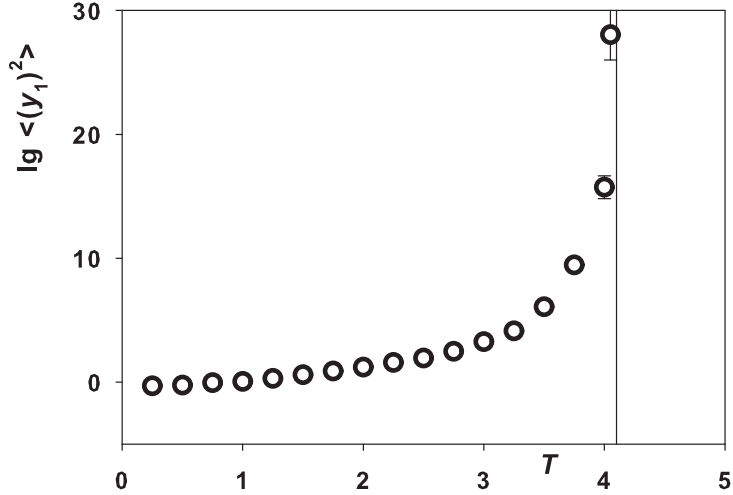}
\end{center}
 \caption{
  \label{fig_8}
  Dependence of mean value $\left< (y_1)^2 \right>$ {\it vs}.
  temperature $T$. At  temperature $T$ close to $T_{\rm thr}
  \approx 4{.}1$ process diverges. Averaging over 20 MD
  trajectories $2\,10^4$ t.u. length each. $\left< (y_1)^2
  \right> \approx 0{.}61$ at $T \to 0$.
  }
\end{figure}

Equations \eqref{2-13} define the stationary random process only
if temperature $T < T_{\rm thr}$. As temperature approaches the
value $T_{\rm thr}$, quadratic mean values diverge. It is shown
in Fig.~\ref{fig_8}. And this is the third method to find $T_{\rm
thr}$.

\begin{figure}
\begin{center}
\includegraphics[width=100mm,angle=0]{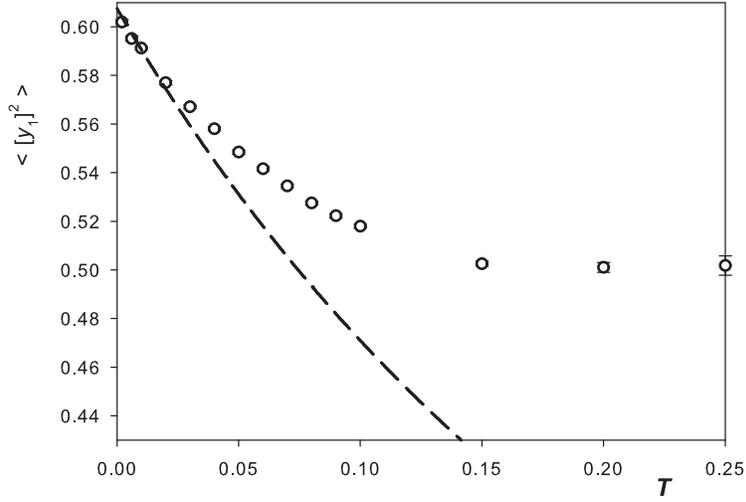}
\end{center}
 \caption{
  \label{fig_9}
  Dependence of the mean squared displacement $\left< [y_1(t)]^2 \right>$
  vs. temperature $T$. Circles -- MD simulation of SDEs
  \eqref{2-12}; solid line -- model of mean rigidities in the harmonic
  approximation. Averaging over 20 trajectories $2 \, 10^4$ t.u.
  each.
  }
\end{figure}

There was analyzed the case of low temperatures $T$ when process
${\bf x}^0$ is  ``weak''. Then rigidity coefficients $g_i$ are
close to unity (see \eqref{2-11}). And as an example we consider
the lattice where actual rigidity coefficients $g_i$ \eqref{2-11}
are substituted by the mean value taken from the equilibrium
Gibbs distribution $\overline g_i=g_0(T)$ and $g_0(T) = 1 +
3\left< (x^0_i -x^0_{i-1})^2 \right>$. This harmonic model is
exactly solvable and results are shown in Fig.~\ref{fig_9}
One can see that process ${\bf y}(t)$ is damped out in the model
with constant rigidity in contrast to the case when actual values
\eqref{2-11} are used. And one can conclude that the growth of
process ${\bf y}(t)$, when temperature increases, is determined
by an {\it increase of fluctuations} but not only by the increase
of rigidities.

\begin{figure}
\begin{center}
\includegraphics[width=100mm,angle=0]{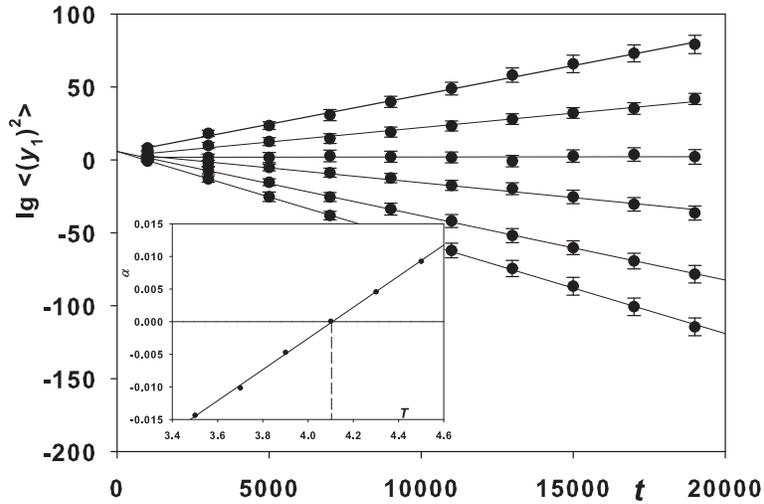}
\end{center}
 \caption{
  \label{fig_10}
  Exponential dependence of the quadratic form $(y_1(t))^2
  \propto \exp (- \alpha t)$ on time. Temperatures from bottom to
  top: $T = 3{.}5$, $3{.}7$, $3{.}9$, $4{.}1$, $4{.}3$, $4{.}5$.
  Averaging over 20 MD trajectories, $2\,10^4$ t.u. each. The
  dependence of coefficient $\alpha$ on temperature is shown in
  insert. $T_{\rm thr} \approx 4{.}1$ is found from the condition
  when $\alpha = 0$.
 }
\end{figure}

Process ${\bf y}(t)$ diverges at high temperatures. And it gives
one more possibility, the fourth one, to find the threshold
temperature. To attain this end the equilibrium process ${\bf
x}^0(t)$ at temperature $T$ is established. Then process ${\bf
y}(t=0)$ is excited in some or other way (its initial conditions
do not influence the final results). And the evolution of the
${\bf y}(t)$ is analyzed. One can see (Fig.~\ref{fig_10}) that
the process exponentially damps out at $T < T_{\rm thr}$ and
exponentially grows at $T > T_{\rm thr}$.

It should be stressed out that the method just described differs
from the previous one (Fig.~\ref{fig_6} and discussion). The
nonlinear case was considered there and its stationarity was
conditioned by nonlinear terms in forces which are absent in the
harmonic approximation.

\begin{figure}
\begin{center}
\includegraphics[width=100mm,angle=0]{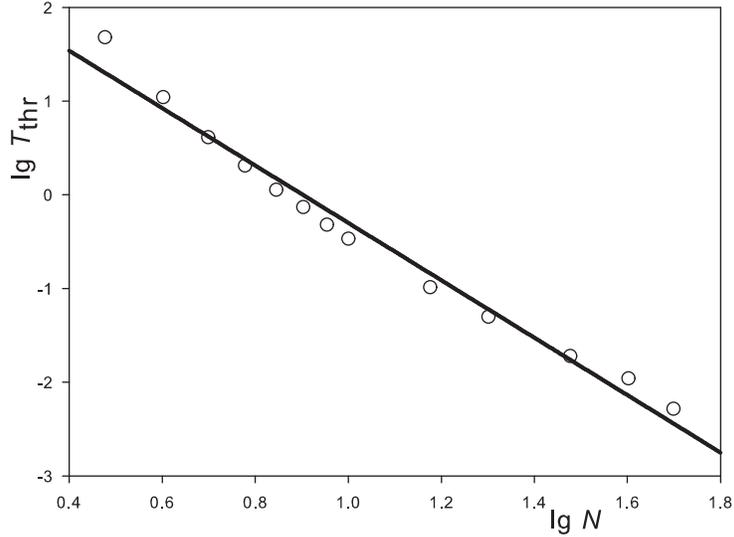}
\end{center}
 \caption{
  \label{fig_11}
  Dependence of $T_{\rm thr}$ vs. lattice length $N$ in log-log
  coordinates. Solid line is the linear fitting $T_{\rm thr} \sim
  N^{-3}$.
 }
\end{figure}

Four methods give the threshold temperature $T_{\rm thr} \approx
4{.}1$. This temperature was found for a fixed lattice length $N
= 5$. Larger lattice lengths were considered and the dependence
of $T_{\rm thr}$ on the lattice length $N$ is shown in
Fig.~\ref{fig_11}. Approximate fitting gives dependence $T_{\rm
thr} \approx 6 \cdot 10^2 \, N^{-3}$.
It means that the majority of usually studied lattices are in the
state when their temperatures are much higher then the threshold
temperature (e.g. if $N > 100$ then $T_{\rm thr} < 6 \cdot
10^{-4}$).


\section{Sound velocity and solitons in $\beta$-FPU lattice}

\begin{figure}
\begin{center}
\includegraphics[width=100mm,angle=0]{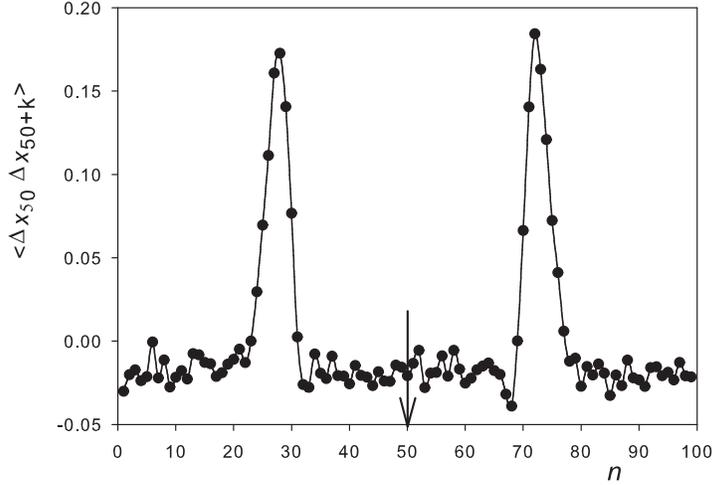}
\end{center}
 \caption{
  \label{fig_12}
Correlator $\left< \Delta x_{50}(t) \, \Delta x_{50+m}(t+20)
\right>$ versus the lattice position (number). $\beta$-FPU
lattice of $N = 101$ oscillators is used. $T=2$. Arrow shows the
site position $n=50$
 }
\end{figure}

Heat conductance is observed in both regimes, -- higher and below
the threshold temperature. And an attempt was undertaken to find
the soliton contribution to the heat conductance at $T > T_{\rm
thr}$. With this in mind, the correlator $\left< \Delta x_k(t) \,
\Delta x_{k+m}(t+\tau) \right>$ was analyzed ($\Delta x_i(t)$ is
the displacement of $i$th oscillator from equilibrium at time
instant $t$). In numerical simulations we fixed the time shift
$\tau = 20$ t.u. and calculated the corresponding correlator ($N
= 101, \ T=2$). Results are shown in Fig.~\ref{fig_12}.
The correlator $\left< \Delta x_{50}(t) \, \Delta x_{50+m}(t+20)
\right>$ has peaks at the coordinate shifts $m = \pm 25$. It
allows to calculate the velocity of excitation propagation
$v_{\rm exc}$ and $v_{\rm exc} \approx 1{.}25$. This velocity is
higher then the sound velocity calculated in the harmonic
approximation $v_{\rm sound}^{\rm harm} =1$ at $\beta = 1$.

Initially these peaks were attributed to solitons. But more
thorough analysis shows that this concepts is not valid. Let we
have the $\beta$-FPU potential $u(y) = \frac12 y^2 + \frac14 y^4$
($\beta = 1$ and $y_i = x_i - x_{i-1}$). In \cite{Ala01} it was
shown that there exists a spectrum of frequencies which are
proportional to the harmonic ones, according to a well defined
law. Therefor the $\beta$-FPU potential can be represented as
\begin{equation}
  \label{f1}
  u(y) = \left( 1 + \frac12 y^2 \right)\frac12 y^2
\end{equation}
and an expression in brackets can be replaced by an effective
harmonic rigidity
\begin{equation}
  \label{f2}
  u(y) = k_{\rm eff}\frac{1}{2} y^2
\end{equation}
The problem is to find $k_{\rm eff}$. It can be done in terms of
a mean field approximation (MFA). Mean value of potential energy
is
\begin{equation}
  \label{f3}
  \left< u_{\rm p}(y) \right> = k_{\rm eff}\frac{1}{2} \left< y^2
  \right>,
\end{equation}
where $\left<  y^2 \right>$ is the mean value of $y^2$.

\begin{figure}
\begin{center}
\includegraphics[width=100mm,angle=0]{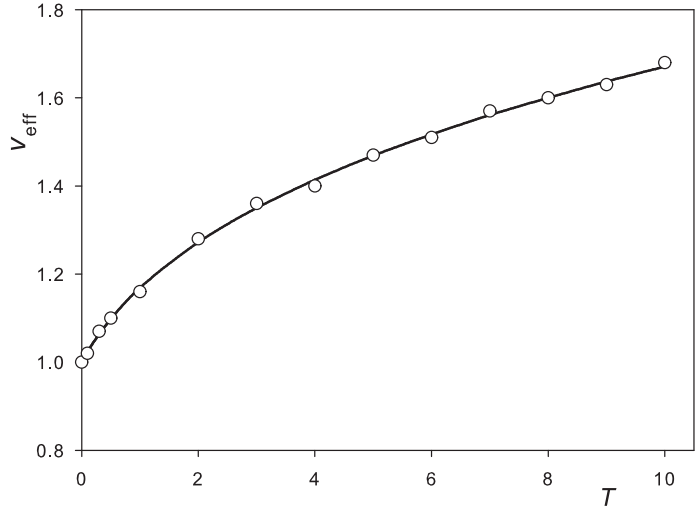}
\end{center}
 \caption{
  \label{fig_13}
The dependence of $v_{\rm eff}$ versus temperature $T$: solid
line -- dependence \eqref{f9}; empty circles -- MD simulation.
 }
\end{figure}

In the harmonic approximation (at not too high temperatures) mean
values of potential and kinetic energies are equal  $\left<
u_{\rm p} \right> = \left< u_{\rm k} \right>$. In canonical
ensemble the identity $\left< u_{\rm k} \right> \equiv T/2$ is
valid for 1D systems. Then
\begin{equation}
  \label{f5}
   k_{\rm eff} \frac12 \left< y^2 \right> = \frac{T}{2}
\end{equation}
The self consistency of the MFA is (expression in brackets in
\eqref{f1} = $k_{\rm eff}$)
\begin{equation}
  \label{f6}
  \left(  1 + \frac12 \left< y^2 \right>  \right) = k_{\rm
  eff}
\end{equation}
From \eqref{f5} it follows that $\left<  y^2 \right> = T/k_{\rm
eff}$ and substitution of $\left<  y^2 \right> = T/k_{\rm
  eff}$ into \eqref{f6} gives the self-consistent equation
  for $k_{\rm   eff}$
\begin{equation}
  \label{f7}
  1 + T/(2 k_{\rm eff}) = k_{\rm   eff}
\end{equation}
with the solution
\begin{equation}
  \label{f8}
  k_{\rm   eff} = \frac12 + \sqrt{\frac14 + \frac{T}{2}}
\end{equation}
Thereby the the effective (``nonlinear'') sound velocity
\begin{equation}
  \label{f9}
v_{\rm eff} = \sqrt{k_{\rm eff}} = \sqrt{\frac12 + \sqrt{\frac14
+ \frac{T}{2}}}, \ \ (m=1)
\end{equation}
is higher then the velocity $v_{\rm sound}^{\rm harm} =1$ found
in the harmonic approximation. Eq.~\ref{f9} gives $v_{\rm eff} =
1{.}27$ for $T = 2$ what coincides with the value $v_{\rm exc}
\approx 1{.}25$ found from correlation functions. The dependence
of effective sound velocity versus temperature is shown in
Fig.~\ref{fig_13}.
Note that the MFA is valid up to very high temperatures $T = 10$,
while this approach originally is well suited only for low
temperatures, and the effective sound velocity exceeds its
harmonic value (at $T = 0$) by $> 50\%$.

\begin{figure}
\begin{center}
\includegraphics[width=140mm,angle=0]{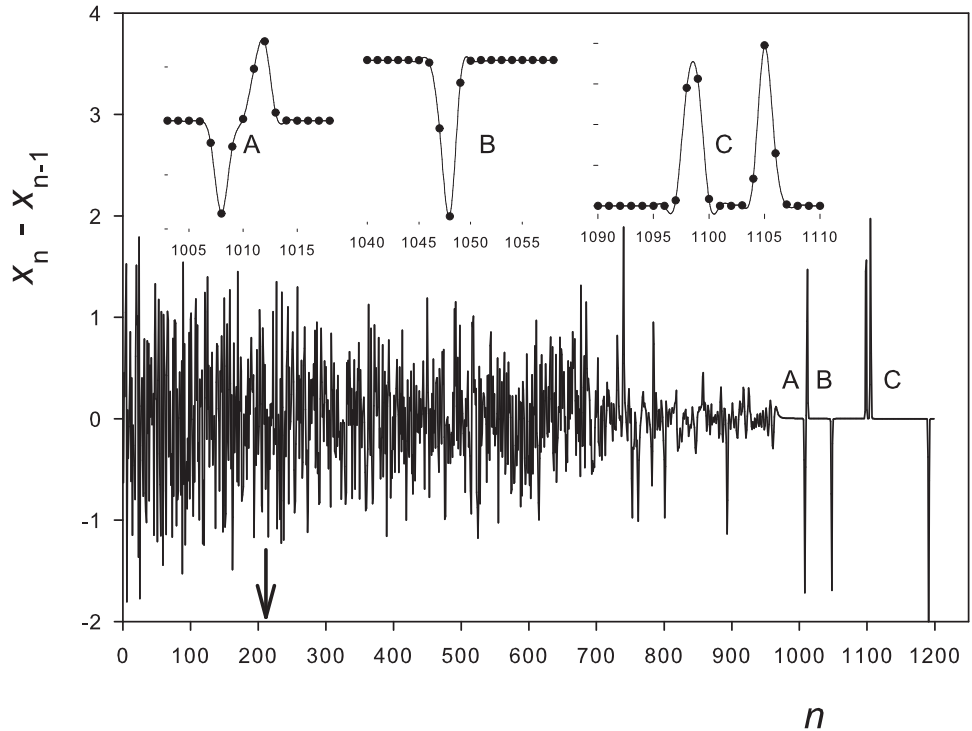}
\end{center}
 \caption{
  \label{fig_14}
Solitons and breasers running out of the lattice. A -- breather,
B --  soliton, C -- pair of antisolitons (solitons of
elongation). Arrow at $n = 200$ shows the border separating
initially thermalized and ``cold'' parts of lattice. Initial
temperature of the left part ($1 \leq n \leq 200$) of the lattice
$T = 10$.
 }
\end{figure}

Next we try to find direct evidences on the solitons
participation in energy transfer. It was done in the following
manner.  Initially lattice of $N=200$ oscillators was thermalized
for some time to reach the thermodynamic equilibrium. Then the
``cold'' lattice (with zero velocities and displacements) with
1000 oscillators was switched to the right end of the lattice.
Solitons, if they exist in the initial lattice, should ``run
out'' to the cold lattice. The same is valid for the moving
breathers. (Note that in the continuum approximation the mKdV
equation corresponds to the discrete $\beta$-FPU potential. And
one can find analytical expressions for solitons of compression,
 antisolitons of elogation and different types of
breathers in \cite{Lam80} ). We waited some time till excitations
run out of the lattice to its cold part where they can be
observed. Results are shown in Fig.~\ref{fig_14}

Analogous approach to visualize breathers in 2D lattice with
three different on-site potentials was utilized in \cite{Bik99}
where initially thermalized lattice was cooled from the borders
and breathers were detected after thermal noise was deleted
through damping boundaries.

The possibility of energy transfer due to solitons was
conjectured three decades ago \cite{Tod79}. Less studied is the
possibility of energy transfer by breathers. One suggested
mechanism is the Targeted Energy Transfer \cite{Kop01, Man04}
when an efficient energy transfer can occur under a precise
condition of nonlinear resonance between discrete breathers.
Various aspects and possible applications of energy transfer by
breathers are considered in \cite{Aub06}.


\section{conclusions}

In conclusion we briefly summarize our main results. A new method
is developed which allows considerable decreasing of the
computation time in calculations of the heat conductance at low
temperature gradients (temperature $T + \Delta T$ of the left
lattice end and $T$ -- of the right end and $\Delta T/T \ll 1$).
This success was achieved by the separation of the total process
of heat conductance into two parts: an equilibrium process ${\bf
x}^0(t)$ at equal temperatures $T$ of both lattice ends and
non-equilibrium process ${\bf x}^1(t)$, responsible for the
energy transport, which occurs at temperature $\Delta T$ of one
end and zero temperature of other end. The equilibrium
(background) process strongly influences the transport
properties: there exists the threshold temperature
$T_{\mathrm{thr}}$ above which some undamped characteristics are
observed; more precisely, correlators of the types $\left<
x^1_i(t) x^1_j(t) \right>$ do not tend to zero at $\Delta T \to
0$, as expected, but have certain nonzero values; at $T <
T_{\mathrm{thr}}$ ``normal'' dependence is observed, i.e. these
correlators have zero values when $\Delta T \to 0$. The reason of
two distinct behaviors is not due to the temperature of the
background process ${\bf x}^0(t)$ but sooner to the temperature
fluctuations. An illustrative example of one variable is briefly
analyzed where the threshold temperature  is also found. The
model of one variable has a rich family of solutions depending on
the parameters and serves to be investigated in more thoroughly.

The threshold temperature was found by few methods and scales
$\sim N^{-3}$ with the lattice size $N$. All practically
interesting systems lies above $T_{\rm thr}$. The threshold
temperature is not sharply pronounced and arbitrarily separates
two mechanisms of the heat conduction: the phonon mechanism
prevails at $T < T_{\rm thr}$, and at $T > T_{\rm thr}$ the
soliton contribution starts to play more significant role with
the increase of temperature. Highly probable that the temperature
fluctuations are responsible for the solitons generation.

Analytical solutions for solitons and breathers are known for the
$\beta$-FPU lattice and these excitations were directly observed
in numerical experiments. Our findings on the soliton
contribution to the heat conductance are in accordance with the
general scenario of heat conductance: phonons gave main
contribution to the heat conductance at low temperatures and
solitons more and more dominate when temperature increases.

We found no relations between the well known weak and strong
stochasticity thresholds and the threshold temperature in the
present paper: they have different energy ranges and different
dependencies on $N$. Additional difference is due to different
statistical ensembles used: traditionally stochasticity
thresholds are found in microcanonical ensemble, but critical
temperature is observed in canonical ensemble where energy
equipartition is realized at any temperature. Statistical
properties do coincide in the thermodynamical limit $N \to
\infty$ for $\mu$-canonical and canonical ensembles, but the
dynamical properties can differ.




\newpage

\appendix

\section{Comparison of efficiencies of different methods
         in computation of heat conductance}
  \label{Accur_comp} 

The main problem in the calculation of heat conductance is to
find the heat current with an appropriate standard error. And we
consider below the effectiveness of the separation of the total
process into the sum of two: ${\bf x}(t) = {\bf x}^0(t) + {\bf
x}^1(t)$ in the sense of computer time expenditure (see
\eqref{2-2} and \eqref{2-6}-\eqref{2-7}\,), where ${\bf x}^0$ is
the equilibrium background process and process ${\bf x}^1$ is
responsible for the heat transport.

\begin{figure}
\begin{center}
\includegraphics[width=90mm,angle=0]{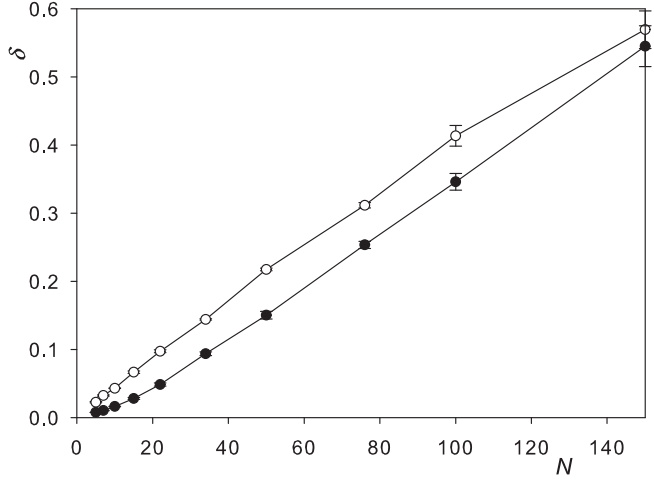}
\end{center}
 \caption{
  \label{fig_15}
  Standard errors of the heat current computed by solving SDEs
  \eqref{2-2} (empty circles) and new SDEs
  \eqref{2-6}-\eqref{2-7} (filled circles). Temperature $T = 0{.}1$,
  $\Delta T = 0{.}01 T$. Averaging over $M =100$ MD trajectories $10^4$ t.u.
  each; total time of computation $10^6$ t.u.
 }
\end{figure}

The comparative efficiency of two approaches (`old' -- solving of
SDEs \eqref{2-2} and `new' -- two SDEs \eqref{2-6}-\eqref{2-7}\,)
to the calculation of the heat flux can be estimated as the
relation of their standard errors $\delta$ at equal conditions of
computation: $\mathrm{Eff} = \delta_{\rm old}/\delta_{\rm new} $.
The result is shown in Fig.~\ref{fig_15}. One can see that the
standard error is systematically less in the suggested approach
as compared to the usually utilized.

\begin{figure}
\begin{center}
\includegraphics[width=90mm,angle=0]{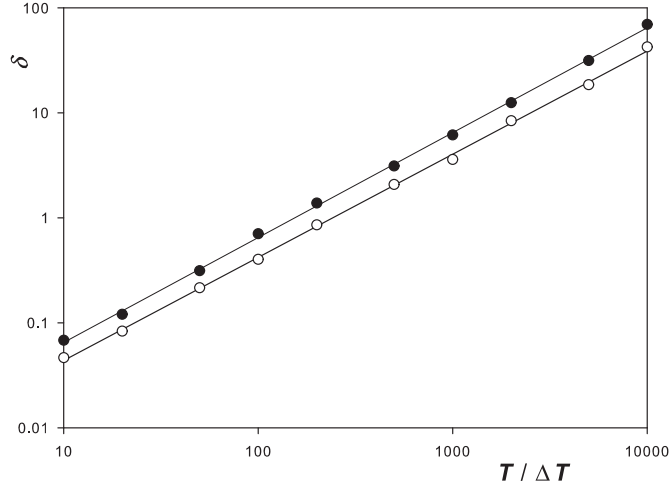}
\end{center}
 \caption{
  \label{fig_16}
  Standard errors vs. inverse temperature gradient $\Delta T$. Filled
  circles -- results of solution SDEs \eqref{2-2} and filled
  circles -- solution of \eqref{2-6}-\eqref{2-7}. $T = 1$, $M =100$
  MD trajectories $10^4$ t.u. length each in both cases.
 }
\end{figure}

More impressive is the behavior of efficiency at decreasing of
the temperature gradient $\Delta T$ (see Fig.~\ref{fig_16}) and
at equal parameters of numerical simulations. Here it should be
emphasized that standard errors increase with the diminishing of
$\Delta T$ in wide range $10^{-4} \le \Delta T/T \le 10^{-1}$ and
the efficiency Eff $> 1$ remains as before. The standard error
$\delta$ increases as $\delta \sim T/\Delta T$.

\section{An example of one variable}
  \label{One_var} 

We consider an example of an equation with one variable $x$ for
the harmonic oscillator with damping
\begin{equation}
  \label{a1}
  \ddot x = - k(t) x - \gamma \dot x \,.
\end{equation}
where $k(t)$ is the stochastic rigidity. This equation is the
illustrative analogue of multi-variable SDEs equations
\eqref{2-7} for the process ${\bf x}^1(t)$. The variable
substitution $x \exp(-\gamma t/2) \to X$ eliminates the damping
and \eqref{a1} can be reduced to
\begin{equation}
  \label{a2}
  \ddot X = - k(t) X  \,.
\end{equation}

Potential energy \eqref{2-11} has the form $u = \frac12 g(t) y^2$
in the case of one variable, where $g(t) = 1 + 3 \, \chi^2(t)$
and $\chi(t)$ -- stochastic process generated by the background
process ${\bf x}^0(t)$.  And it is reasonable to choose the
random rigidity in \eqref{a2} in the form
\begin{equation}
  \label{a3}
  k(t) = 1 + \varepsilon z^2(t) \,,
\end{equation}
where $\varepsilon$ is free parameter and $z(t)$ is the
stationary random process describing the dynamic of the harmonic
oscillator influenced by Langevin source with temperature $T$:
\begin{equation}
  \label{a4}
 \ddot z = -z + \xi - \gamma \dot z
\end{equation}
and $\left( \left< \xi(t_1) \, \xi(t_2 \right> = 2 \gamma T \,
\delta(t_1 - t_2) \right)$

We compare the solution $X(t)$ \eqref{a1}--\eqref{a3} with the
solution of well known {\it deterministic} Mathieu equation
\begin{equation}
  \label{a5}
 \ddot y = -[1 + g \cos^2(t)]\, y  .
\end{equation}

Different types of solutions of the Mathieu equation depend on
the parameter $g$ and initial conditions. There exists such
$g_{\rm cr}$ that the solution is the sum of periodic functions
at $g < g_{\rm cr}$, and the solution is the superposition of
periodic functions multiplied by the exponentially increasing and
decreasing function $\exp(\pm \mu t)$ at $g > g_{\rm cr}$.

Equations \eqref{a2}--\eqref{a4} also have a rich family of
solutions depending on initial conditions and parameter values.
As an illustrative example we consider the following set of
parameters: $X(t=0) = 0{.}5, \ {\dot X}(t=0) = 0, \ \varepsilon =
50, \ \gamma = 1$.

\begin{figure}
  \begin{center}
    \includegraphics[width=80mm,angle=0]{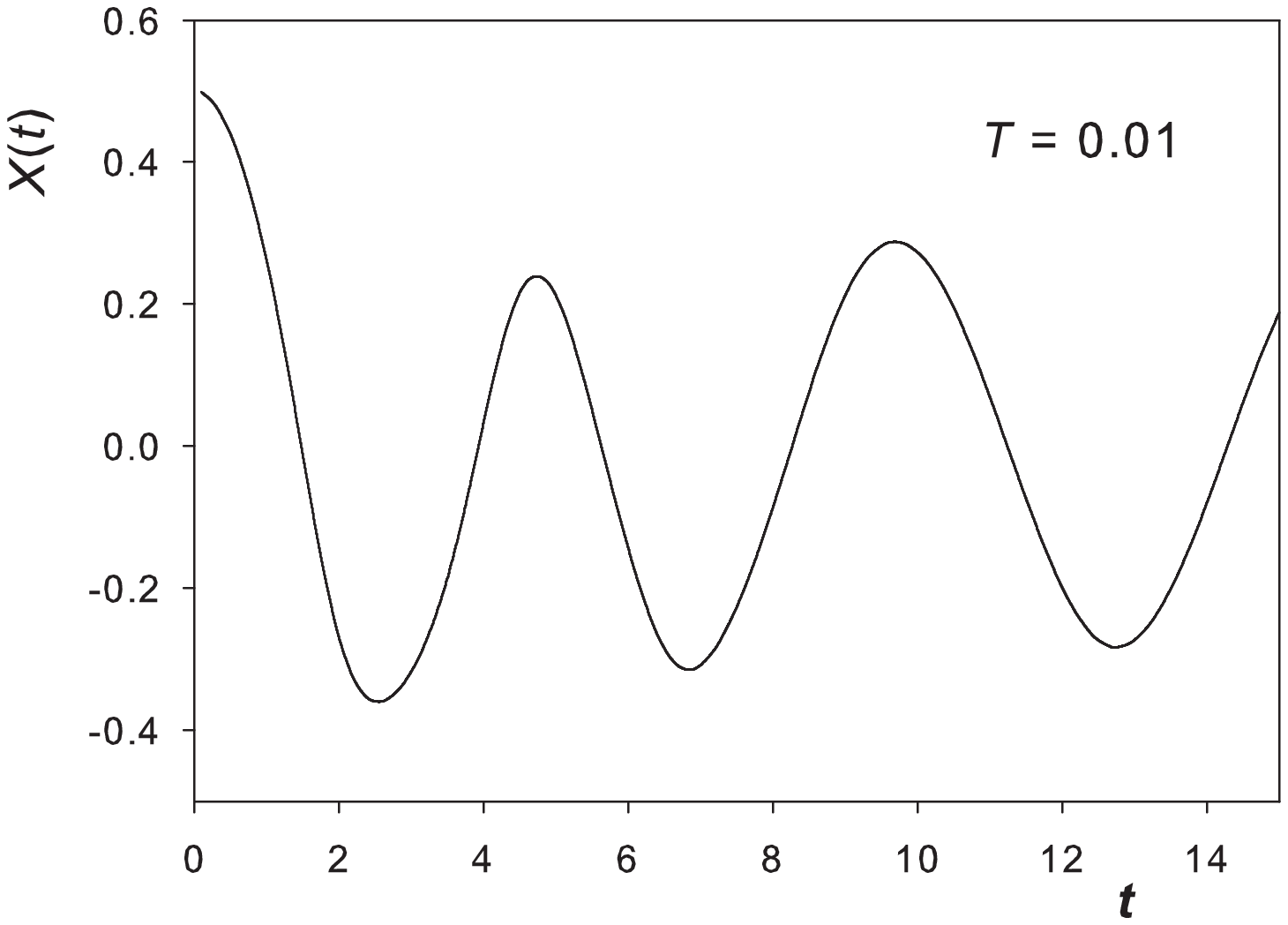}
    \includegraphics[width=80mm,angle=0]{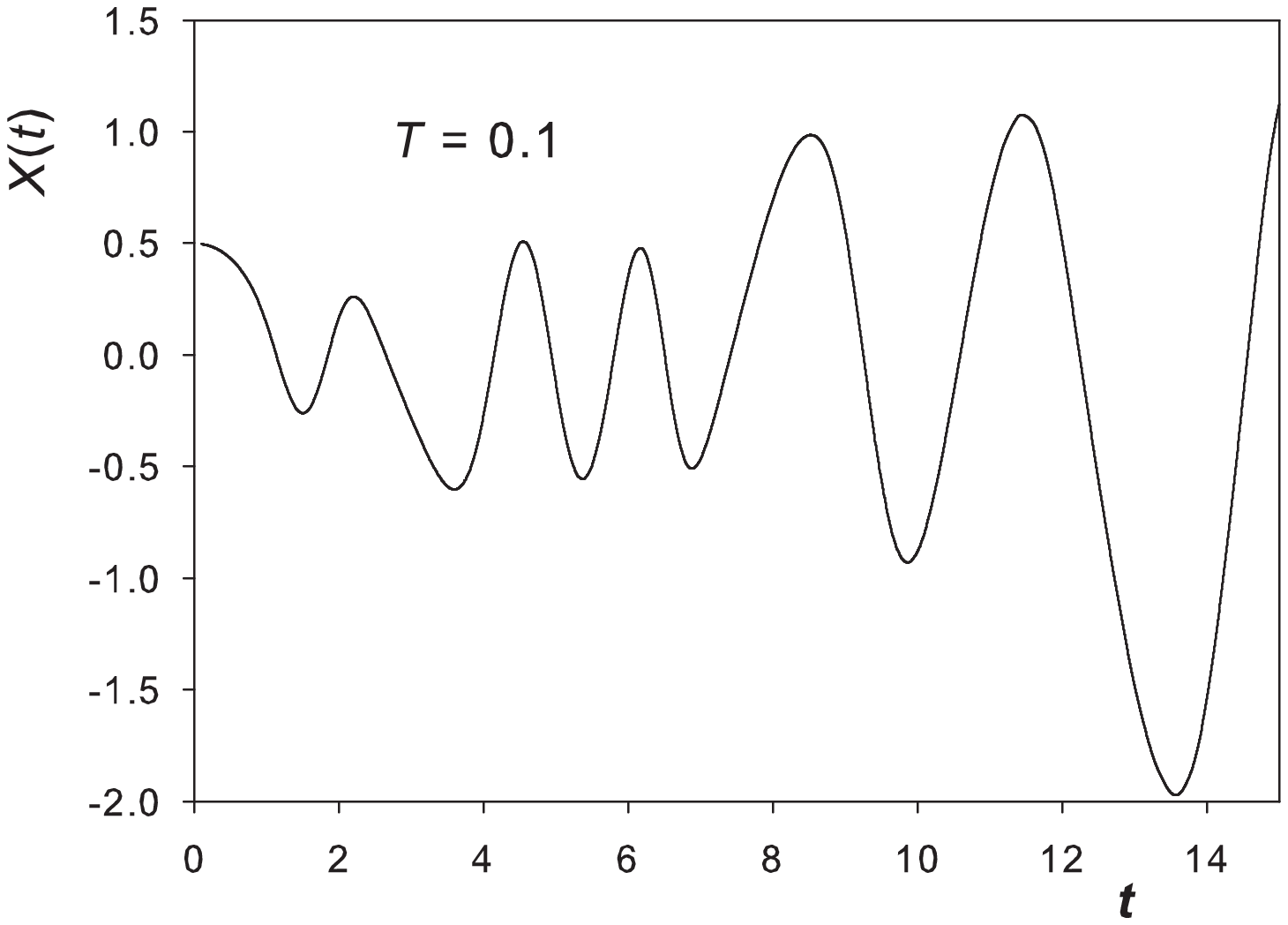}

      \vspace{-0.5 cm}
   a) \hspace{8 cm} b)

   \includegraphics[width=80mm,angle=0]{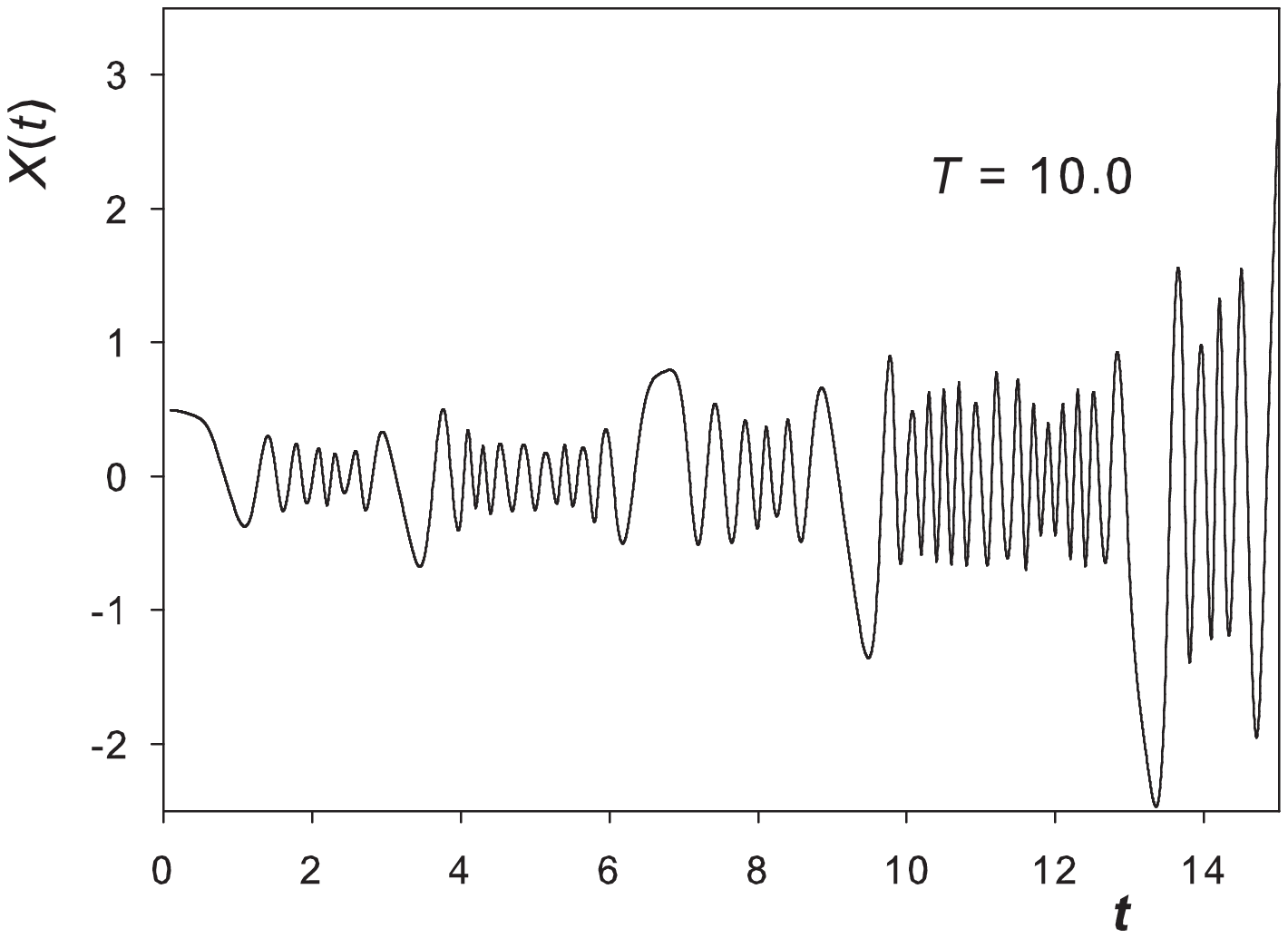}
      \vspace{-0.7 cm} \begin{center}  c) \end{center}

 \end{center}
 \caption{
  \label{fig_17}
 Temporal behavior of process $X(t)$ at different temperatures:
 $T = 0{.}01, \ T = 0{.}1, \ T = 10$. }
\end{figure}

Below we demonstrate only the qualitative behavior of process
$X(t)$ depending on the temperature $T$ of stochastic process
\eqref{a4} and results are shown in Figs.~\ref{fig_17}(a--c). One
can see different regimes as $T$ increases. And there exists some
critical temperature $T_{\rm cr}$ above which the process $X(t)$
diverges $\left( T_{\rm cr} \approx 100 \right)$. It should be
noted that the overall scenario strongly depends on the choice of
initial conditions ($X(t=0), \ \dot X(t=0)$) and the particular
sequence of random Langevin forces $\{ \xi \}$ in \eqref{a4}. At
larger times process $X(t)$ becomes more complex. The full
analysis of system \eqref{a2}--\eqref{a4} is not our primary
goal, but these equations serve more intensive attention.

\end{document}